\documentclass{aa}  
\usepackage{longtable}
\usepackage{graphicx}
\usepackage{txfonts}
\usepackage{subcaption}         % necessary for continued figures, example in section 3
\usepackage{lscape}             % to rotate a single page table, example in appendix.
\usepackage{pdflscape}                                % For landscape tables, see the longtable examples.
\usepackage{placeins}           % useful with \FloatBarrier, to keep 
\usepackage{tabularx}                                % onecolumn floats from drifting to the next section
\usepackage{newtxtext,newtxmath}
\usepackage{pdflscape}
\usepackage[T1]{fontenc}
\usepackage{ulem}
\DeclareRobustCommand{\VAN}[3]{#2}
\let\VANthebibliography\thebibliography
\def\thebibliography{\DeclareRobustCommand{\VAN}[3]{##3}\VANthebibliography}
\usepackage{graphicx}	% Including figure files
\usepackage{amsmath}	% Advanced maths commands
\usepackage{xfrac,nicefrac}

\newcommand{\Mdot}{\mbox{\,$\rm M_{\odot}$}}        % solar mass
\newcommand{\Zdot}{\mbox{\,$\rm Z_{\odot}$}}        % solar metallicity
\newcommand{\ZSMC}{\mbox{\,$\rm Z_{\rm SMC}$}}        % SMC metallicity
\newcommand{\Ldot}{\mbox{\,$\rm L_{\odot}$}  }        % solar luminosity
\newcommand{\aov}{$\alpha_{\rm ov}$}		% overshoot alpha
\newcommand{\fov}{$f_{\rm ov}$}		% overshoot alpha
\newcommand{\amlt}{$\alpha_{\rm mlt}$}		% mlt alpha
\begin{document}
% \nolinenumbers
\nolinenumbers
\title{The impact of wind mass loss on nucleosynthesis and yields of very massive stars at low metallicity}
\titlerunning{Low Z yields}

\author{
Erin R. Higgins\inst{1,2} \and 
Jorick S. Vink\inst{1}\and 
Raphael Hirschi\inst{3,4}\and 
Alison M. Laird\inst{5} \and 
Gautham Sabhahit\inst{1}}
\institute{Armagh Observatory and Planetarium, College Hill, Armagh BT61 9DG, N. Ireland 
\and Astrophysics Research Centre, School of Mathematics and Physics, Queen's University Belfast, Belfast, BT7 1NN, UK
\and Astrophysics Group, Keele University, Keele, Staffordshire ST5 5BG, UK
\and Kavli Institute for the Physics and Mathematics of the Universe (WPI),\\ University of Tokyo, 5-1-5 Kashiwanoha, Kashiwa 277-8583, Japan
\and School of Physics, Engineering and Technology, University of York, York, YO10 5DD\\
\email{e.higgins@qub.ac.uk}}

\date{Accepted 15 April 2025. Received 16 December 2024.}

\abstract{
The chemical feedback from stellar winds in low metallicity (Z) environments is key for understanding the evolution of globular clusters and the early Universe. With disproportionate mass lost from the most massive stars (M$>$ 100\Mdot), and an excess of such stars expected at the lowest metallicities, their contribution to the enrichment of the early pristine clusters could be significant. In this work, we examine the effect of mass loss at low metallicity on the nucleosynthesis and wind yields of (very) massive stars. We calculate stellar models with initial masses ranging from 30 to 500\Mdot\ during core Hydrogen and Helium burning phases, at four metallicities ranging from 20\% \Zdot\ down to 1\% \Zdot. The ejected masses and net yields are provided for each grid of models. While mass-loss rates decrease with Z, we find that not only are wind yields significant, but the nucleosynthesis is also altered due to the change in central temperatures and therefore also plays a role. We find that 80-300\Mdot\ models can produce large quantities of Na-rich and O-poor material, relevant for the observed Na-O anti-correlation in globular clusters.
}

\maketitle
\section{Introduction}
Mass loss from very massive stars (VMS) with masses over 100 \Mdot\ at low metallicity is a key factor in shaping their evolution and chemical yields. Unlike stars at solar metallicity, where line-driven winds dominate the mass loss, stars in low-$Z$ environments experience significantly weaker winds \citep{Vink01}. This leads to less efficient angular momentum loss and a potentially more prolonged phase of rapid rotation, which can enhance internal mixing and alter nucleosynthesis outcomes. Studies suggest that VMS at low metallicity could retain much of their mass until late evolutionary stages, resulting in distinct chemical feedback patterns compared to their higher-$Z$ counterparts \citep{yusof13}.

The chemical evolution of globular clusters (GCs) has long been a subject of intense study, particularly due to the well-established abundance anomalies such as the Na-O anti-correlation \citep{carretta09,charbonnel16,bastianlardo18}. These anomalies suggest that GCs experienced self-enrichment processes early in their formation, likely driven by feedback from a first population of stars. Understanding the sources of this chemical enrichment requires detailed models of stellar nucleosynthesis and mass loss, particularly for stars at low metallicity ($Z$). The focus on low-$Z$ environments is motivated by their relevance to both the early Universe and nearby dwarf galaxies which are considered analogs of the conditions in the first galaxies.

The initial mass function (IMF) in these low-metallicity environments is thought to be more top-heavy, with a higher proportion of very massive stars than is typically seen in the present-day Universe. This expectation follows from theoretical studies of star formation under primordial conditions \citep{Bromm2004} and diminished wind feedback \cite{vink18}. First empirical evidence for a more top-heavy IMF at LMC metallicity ($\sim$50\% solar $Z$) was presented in \cite{Schneider18}. The feedback from these VMS is considered to play a crucial role in driving early galaxy evolution, providing both chemical enrichment and energy to their surroundings \citep{krause13,vinkbook,crow16}. In dwarf galaxies, the presence of abundance anomalies similar to those found in GCs points to a similar population of massive stars contributing to the chemical evolution \citep{Gratt04}.

Previous works have focused on the nucleosynthetic yields of massive stars, particularly at low metallicity, and their potential to explain the observed abundance patterns in both GCs and the broader Universe. These studies have emphasized the role of rotational mixing at low $Z$, which may potentially be able to bring products of nuclear burning to the surface, thereby contributing to the observed Na-O anti-correlation \citep{Decressin07}. It should however be noted that such a model relies not only on high stellar rotation rates, but also very efficient rotational mixing, which has been questioned \citep{vink-harries17}. 
\cite{Denissenkov} proposed supermassive stars (M$_{*} > 10^{4}$ \Mdot) as the source for observed anti-correlations in GCs due to losing significant amounts of mass within the first 10\% of core H-burning. However, \cite{Denissenkov} also suggest that super-Eddington winds would be required to provide such high mass loss on this timescale, crucially before further nucleosynthesis.

Instead, the early Universe was likely populated by a significant number of VMS with masses in the range 100-1000\Mdot, which may have contributed to the reionization and chemical enrichment of the intergalactic medium \citep{Schaerer24}. These stars are thought to have provided the first significant chemical feedback in the Universe, enriching primordial gas and setting the stage for subsequent generations of star formation \citep{Baraffe2001}. In this context, studying the yields of very massive stars at low metallicity not only helps us understand the formation of globular clusters but also provides insights into the broader processes that shaped the early Universe.

This paper aims to investigate the wind yields from very massive stars at low metallicity, focusing on their contributions to chemical feedback in globular clusters and dwarf galaxies. We explore how different mass-loss rates, rotation speeds, and metallicities influence the chemical enrichment patterns, particularly in the context of Na-O anti-correlations and other abundance anomalies. By comparing our theoretical yields with recent observations, we aim to provide new insights into the role of VMS in the early chemical evolution of galaxies.

\begin{table}
    \centering
        \caption{Initial composition of our model grid for a range of initial Z in mass fractions.}

    \begin{tabular}{c c c c}
    \hline
Z	&Y	&X\\
      \hline \hline

0.004	&0.248	&0.748\\
0.002	&0.244	&0.754\\
0.00067&	0.24134	&0.75799\\
0.0002	&0.2404	&0.7594\\
        \hline
    \end{tabular}
    \label{tab:initZ}
\end{table}

 \begin{figure}
    \centering
    \includegraphics[width = \columnwidth]{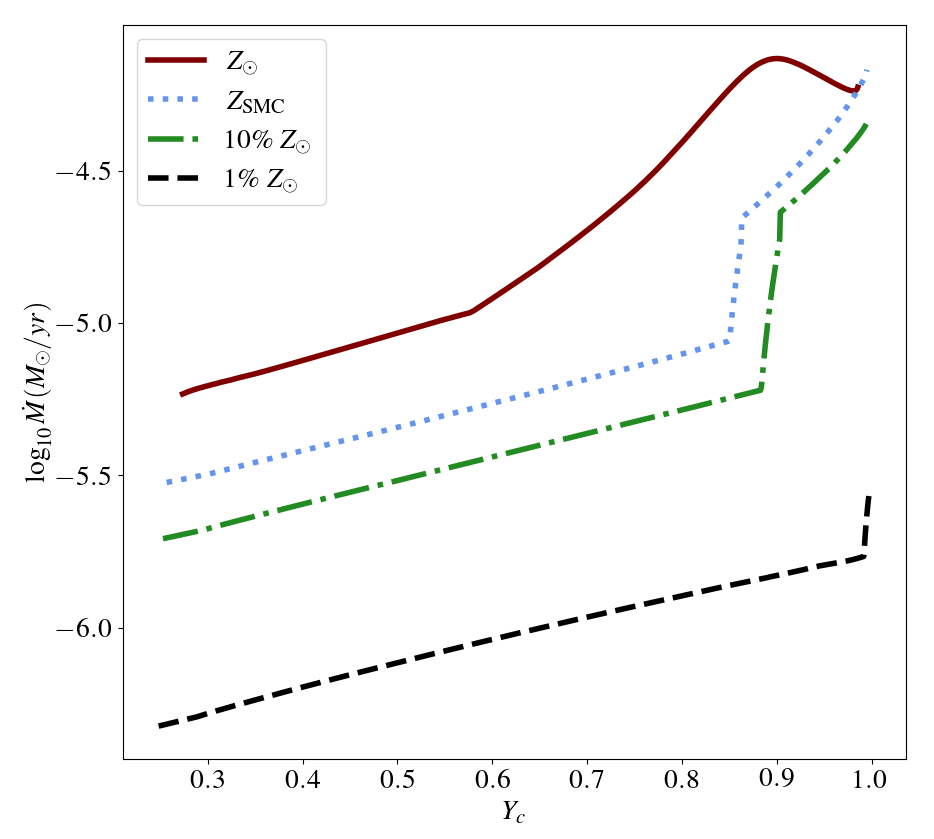}
    \caption{Mass-loss rates of 100\Mdot\ models for each Z as a function of core He abundance during the MS evolution.}
    \label{fig:mdot_MS_allZ}
\end{figure}
\section{Stellar models}\label{masslossrotation}
\begin{figure*}
    \centering
    \includegraphics[width = \textwidth]{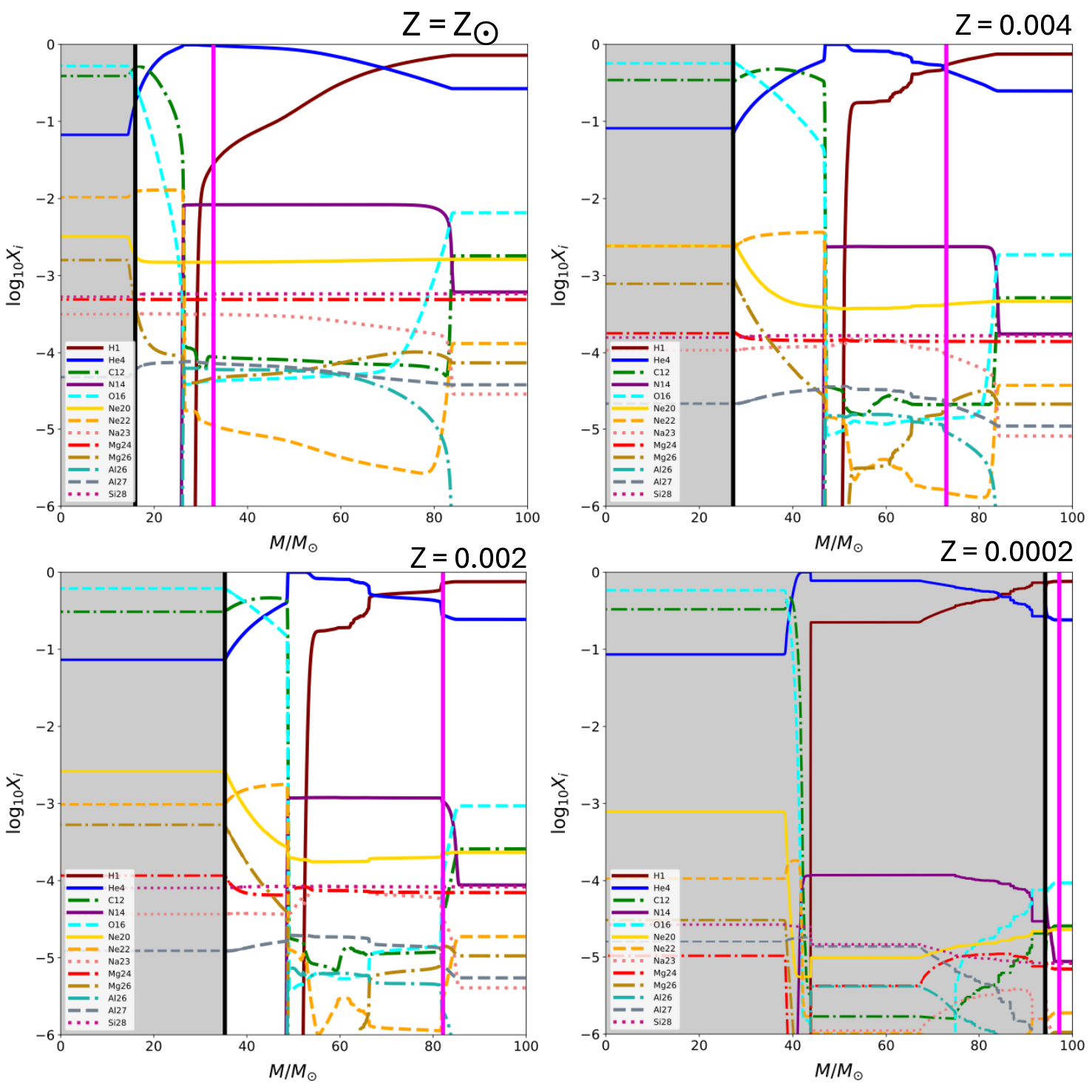}
    \caption{Overview of the chemical composition and its evolution for a 100\Mdot\ model, calculated from the onset of core H-burning until core He-exhaustion, for \Zdot\ (upper left), Z$=$ 0.004 (upper right), Z$=$ 0.002 (lower left) and Z$=$ 0.0002 (lower right). The shaded region illustrates the interior composition at the end of core He-burning, while the unshaded region shows the surface abundances of each isotope with time evolving from right to left as mass is lost through winds. The vertical black solid line denotes the end of core He-burning, while the vertical magenta solid line shows core H-exhaustion.}
    \label{fig:yield100Z}
\end{figure*}

We calculate grids of models using the one-dimensional stellar evolution code \texttt{MESA} \citep[r10398;][]{Pax11,Pax13,Pax15,Pax18,Pax19} for a grid of initial masses of 30, 50, 80, 100, 200, 300, 400, and 500\Mdot. We implement a nuclear reaction network which includes the relevant isotopes for massive star evolution until the end of core O-burning, such that the nucleosynthesis is comparable to \cite{Higgins+23}. This nuclear network comprises the following 92 isotopes: n, $^{1, 2}$H, $^{3, 4}$He, $^{6, 7}$Li, $^{7, 9, 10}$Be, $^{8, 10, 11}$B, $^{12, 13}$C, $^{13-16}$N, $^{14-19}$O, $^{17-20}$F, $^{18-23}$Ne, $^{21-24}$Na, $^{23-27}$Mg, $^{25-28}$Al, $^{27-33}$Si, $^{30-34}$P, $^{31-37}$S, $^{35-38}$Cl, $^{35-41}$Ar, $^{39-44}$K, and $^{39-44,46, 48}$Ca. Our stellar models are computed for a range of metallicities, ranging from \ZSMC\ to 1\% \Zdot\ where \Zdot\ $=$ 0.02, $X$ $=$ 0.720, and $Y$ $=$ 0.26, where the relative composition is adopted from \cite{asplund09}. At each metallicity we calculate Y $=$ Y$_{\rm{prim}}$ $+$ ($\Delta$Y/$\Delta$Z)Z where $\Delta$Y/$\Delta$Z $=$ 2 and Y$_{\rm{prim}}$ $=$ 0.24 \citep{AG89, pols95}, see Table \ref{tab:initZ}. We compare with the \Zdot\ models of \cite{Higgins+23} in Figs. \ref{fig:yield100Z} and \ref{fig:abundances}, and note that in that case models were calculated with \Zdot\ $=$ 0.014. We avail of the OPAL opacity tables from \cite{RogersNayfonov02}, and adopt nuclear reaction rates from the JINA Reaclib Database v2.2 \citep{Cyburt10}. The mixing-length-theory (MLT) of convection describes the treatment of convection in our models, where we apply an efficiency of \amlt $=$ 1.67 \citep{arnett19}. The Schwarzschild criterion defines the convective boundaries in our models and as such we do not implement semiconvective mixing. For convective boundary mixing (CBM), we include the exponential decaying diffusive model of \citet{Freytag1996} \citep[see also][]{Herwig2000}
with \fov $=$ 0.03 (corresponding to \aov $\simeq$ 0.3) for the top of convective cores and shells, and with \fov $=$ 0.006 for the bottom of convective shells. \cite{bowman20review} find a range of \aov\ from asteroseismology results with \aov\ up to 0.4, so our value for the top of convective zones falls in the range of asteroseismology-inferred values. This value also falls in between the majority of published large grids of stellar models such as $\alpha_{\textrm{ov}}=0.1$ in \citet[][]{Ekstroem2012}, $\alpha_{\textrm{ov}}$ $=$ 0.335 in \citet[][]{Brott2011}, and recent studies on CBM \citep[][]{Higgins, Scott21} supporting values for \aov\ up to at least 0.5 for stars above 20\,M$_\odot$. For the bottom boundary, a CBM value of 1/5 the value of the top boundary is based on 3D hydrodynamic simulations \citep{Cristini2017,Cristini2019,Rizzuti2022} finding that CBM is slower at the bottom boundaries due to being stiffer and therefore harder to penetrate.

We apply convection in superadiabatic layers via the \texttt{MLT++} prescription which aids convergence of such models. The temporal resolution of our models has been set with \texttt{varcontrol}\texttt{target} $=$ 0.0001, and a corresponding spatial resolution of \texttt{mesh}\texttt{delta} $=$ 0.5. All calculations begin with a pre-main sequence and then evolve from the ZAMS until core H-depletion ($^{1}\rm{H}_{\rm{c}}$ $<$ 0.0001) or core He-depletion ($^{4}\rm{He}_{\rm{c}}$ $<$ 0.0001). At Z $<$10\% \Zdot\ stellar models evolve at higher temperatures and luminosities, and as such are compact and evolve close to their Eddington limit. Therefore, we calculate the core H-burning MS only for the grids of models with Z $<$10\% \Zdot\ (i.e. Z$=$ 0.00067 and 0.0002), while the higher Z models evolve until core He-exhaustion (10\% and 20\% \Zdot\ grids).

The effect of rotation on nucleosynthesis and wind yields are tested for models with Z $=$ 0.004 and 0.002. We focus on non-rotating models at the lowest Z (1/30th \Zdot\ and 1/100th \Zdot) since these stars can spin up to reach criticality. However, we provide a subset of models initially rotating at 40\% of critical, calculated at 1/30th \Zdot\ to compare with observed Na and O abundance patterns (see Sect. \ref{discussion_dilute}). To explore the consequences of rotation on wind yields at Z $=$ 0.004 and 0.002, we calculate 3 sets of models for each of these Z: non-rotating, $\Omega_{\rm{ini}}/\Omega_{\rm{crit}}$ $=$ 40\%, and 70\%. We adopt rotation with instabilities employed for angular momentum transfer and chemical mixing as described by \cite{Heger00}, and do not include the effects of a magnetic field. 

The wind rates applied throughout our grids of models are invoked for each evolutionary phase and based on the wind driving mechanism at each stage. Therefore, for hot (log$_{10}$ (T$_{\rm{eff}}$/K) $\geq$ 4.0), optically thin OB stars, evolving comfortably below the transition point ($M$ $\lesssim$ 60\Mdot), we adopt the \cite{Vink01} rates. 

We apply the mass-loss framework of \cite{sabh23} where stars switch from optically-thin to optically-thick rates above the transition point. This is calculated iteratively for $L$ and $M$, where the condition for switching to an optically-thick wind is established as a function of the wind efficiency parameter $\eta$. Above the switch, or transition point, we then apply wind rates based on the relation by \cite{vink11}. 

From 4kK $<$ ($T_\mathrm{eff}$/K) $<$ 100kK, we employ the same mass-loss framework on the post-MS as during the MS since the core evolution should not dictate the wind driving mechanism. However, below $T_\mathrm{eff} \approx 4$ kK we switch to the cool supergiant prescription with rates adopted from \cite{deJager}. Finally, for classical Wolf-Rayet stars, where $T_\mathrm{eff} > 100$ kK and $X_\mathrm{c}$ $<$ 0.01, we apply the wind relation from the hydrodynamically-consistent models of \citet{SV2020}. 

For rotating models, we assume that near the radiative Eddington limit $\Gamma > 0.639$, the $\Omega\Gamma$-limit can be reached, and increased mass can be lost, accounting for the von Zeipel theorem. The implementation of this mass-loss relation follows the approximation as derived by \cite{MM00}, 

\begin{equation}
\begin{array}{c@{\qquad}c}
\dfrac{\dot{M}(\Omega)}{\dot{M}(\Omega = 0)} \approx \dfrac{(1-\Gamma)^{\frac{1}{\alpha}-1}}{\Big(1-\frac{4}{9} (\frac{\varv}{\varv_\mathrm{crit, 1}})^2 -\Gamma\Big)^{\frac{1}{\alpha}-1}}
\end{array}
\label{eq: rotation_mass_loss}
\end{equation}
where $\alpha$ is the force multiplier parameter, the critical velocity $\varv_\mathrm{crit,1} = \Big(\frac{2}{3}\frac{GM}{R_\mathrm{pb}}\Big)^{0.5}$, and $R_\mathrm{pb}$ is the polar radius at break-up, where we assume that the polar radius R$_\mathrm{p}$ does not alter with rotation rate ($\Omega$), $R_\mathrm{pb}/R_\mathrm{p}(\Omega) = 1$.

\section{Nucleosynthesis}
 \begin{figure*}
    \centering
    \includegraphics[width = \textwidth]{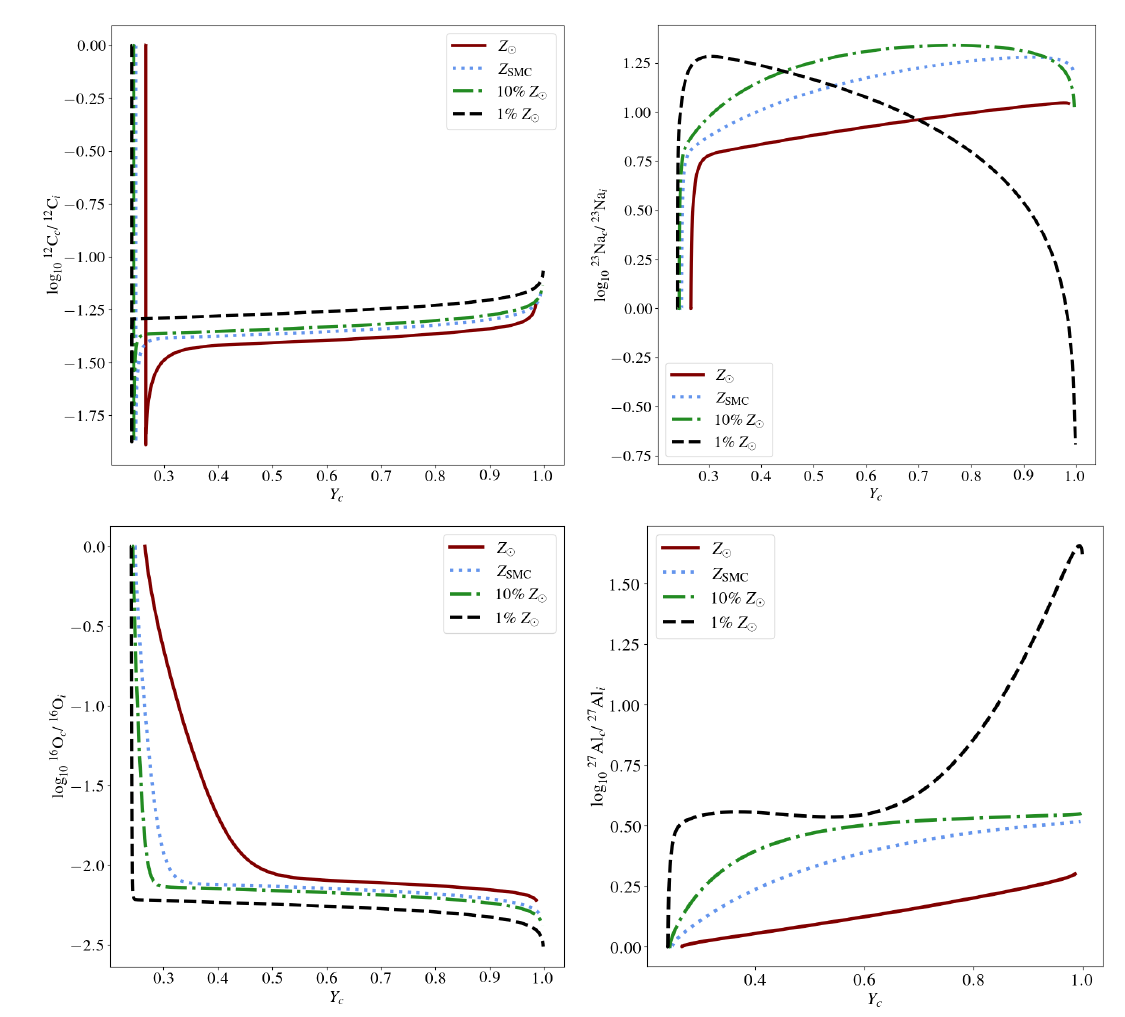}
    \caption{Central abundances of $^{12}$C (upper left), $^{16}$O (lower left), $^{23}$Na (upper right) and $^{27}$Al (lower right) relative to their initial abundance, for each Z as a function of core He abundance during the MS evolution of 100\Mdot\ models.}
    \label{fig:abundances}
\end{figure*}
We present net wind yields and ejected masses for each Z grid of stellar models. Since stellar models with Z $\geq$ 0.1\Zdot\ include rotation,
we can study the effects of rotation on the ejecta at these metallicities. Thus we compile a table of outputs for each Z, with the \ZSMC\ and 0.1\Zdot\ grids including 3 subgrids with different rotation rates (0, 40\% and 70\% critically rotating).  In line with \citet{hirschi05} and \citet{Higgins+23}, the net wind yield calculated for a star of initial mass, $m$, and isotope, $i$, is:
\begin{equation}\label{yieldeq}
    m^{\rm wind}_{i} = \int_{0}^{\tau(m)} \dot{M}(m, t) \,[{X}^{S}_{i}(m, t) - {X}^0_{i}] \,dt 
\end{equation}
where $\dot{M}$ is the mass-loss rate, ${X}^{S}_{i}$ is the surface abundance of a given isotope, and ${X}^0_{i}$ is the initial abundance of a given isotope at the ZAMS. Net yields are then integrated from the beginning of core H-burning until $\tau(m)$, corresponding to the end of core He-burning for \ZSMC\ and 0.1\Zdot\ grids and the end of core H-burning for 1/30th\Zdot\ and 0.01\Zdot\ grids, respectively. 

We also calculate ejected masses, $EM$ of each isotope, $i$, by:
\begin{equation}\label{EMeq}
    EM_{im} = \int_{0}^{\tau(m)} \dot{M} \, {X}^{S}_{i}(m, t) \,dt .
\end{equation}

The key reactions which dominate the wind yields of massive stars involve the CNO cycle, and secondary cycles (including the Ne-Na and Mg-Al cycles) during H-burning which affect abundant isotopes of C, N, O, F, Ne, Na, Mg and Al. During core He-burning, the H-synthesised $^{4}$He produces $^{12}$C via the triple-$\alpha$ reaction, resulting in an increase in $^{12}$C which activates the $^{12}$C($\alpha$, $\gamma$)$^{16}$O reaction. The resulting CO core at core He-exhaustion subsequently can be used to estimate the explodability of the stellar core \citep{oconnor,Farmer+2019}.

At lower Z, the evolution of massive stars is hotter and more compact, relative to higher Z environments. Furthermore, the amount of mass lost during the core H-burning phase decreases as a result of the Z-dependence on radiation-driven winds (see Fig. \ref{fig:mdot_MS_allZ}). In fact, the transition point at which massive and very massive stars switch from an optically thin to optically thick wind also shifts to higher initial masses at lower Z \citep{sabh23}. This means that $\sim$ 80-200\Mdot\ stars will have optically thin winds at Z$\sim$0.1-0.01 \Zdot, and could result in more He-burning products in the ejecta rather than H-rich ejecta.

We explore the amount of mass lost in the H-burning phase compared to the He-burning phase, and what remnant mass is left for a range of initial masses at \ZSMC\ and 0.1\Zdot. As expected, the higher Z models with higher initial masses lose a significant proportion of their total mass already on the MS, with very little mass lost during the post-MS. Similarly, the remnant masses of the 0.1\Zdot\ models are significantly higher than \ZSMC\ models with the same initial mass, because (i) the transition point has shifted from $\sim$100-200\Mdot\ to $\sim$200-300\Mdot, and (ii) stars in lower Z environments lose less mass at all evolutionary stages. Interestingly, the same behaviour is seen from the He-burning $\Delta$ M at each Z and rotation rate, where lower mass stars lose more mass on the post-MS. With increasing initial mass, the post-MS winds become less relevant in comparison to both the MS winds and the remnant masses. 

A key question bridging stellar winds and galactic chemical evolution is whether stellar yields are affected mainly by the Z-dependence on stellar winds, or whether the nucleosynthesis is also affected significantly to alter the ejecta. Figure \ref{fig:yield100Z} shows the surface evolution of isotopes (right to left, white region) and interior composition (grey shaded region) at core He-exhaustion, for a 100\Mdot\ star at \Zdot, \ZSMC, 10\%\Zdot\ and 1\%\Zdot. Firstly, the dramatic change in the stellar mass at core He-exhaustion is seen (grey region, marked by black line), also emphasising the increased ejected mass with increasing Z (white region). For instance, while the \Zdot\ model ejects $\sim$20\Mdot\ more than the 10\%\Zdot\ model, the 10\%\Zdot\ model loses $\sim$ 10 times more mass than the 1\%\Zdot\ model. 

Secondly, while the most abundant elements (H, He, C) are not altered in central mass fraction by Z, the less abundant elements such as Ne, Na, Mg and Al are shown to have different behaviours at lower Z. For instance the amount of N lost on the MS at \Zdot\ is higher than at \ZSMC\ leaving less N to be synthesised into $^{22}$Ne in the post-MS. We can see the change in $^{22}$Ne (orange dashed line) from the top panel to the lower panels. Of course, with decreased Z$_{\rm{Total}}$ each element will have a slightly lower initial abundance for each low Z environment, and different evolutionary timescales for each burning phase. However, by comparing the relative change in each isotope in the core (relative to their initial abundances) during the core H-burning phase as a function of increasing core He abundance (or MS evolution) we find that the abundances can change by $\sim$ 0.3 dex for $^{12}$C and $^{16}$O, and up to $\sim$ 1.5 dex for $^{23}$Na and $^{27}$Al, when comparing \Zdot\ models to low Z models (Fig. \ref{fig:abundances}). 

Finally, we examine the central abundance of $^{12}$C, $^{16}$O, $^{23}$Na and $^{27}$Al relative to their initial composition throughout the MS evolution ($Y_{c}$) in Fig. \ref{fig:abundances}. We find that $^{12}$C and $^{16}$O show central abundance patterns which are not significantly affected by $Z$ during the MS, with only $\sim$ 0.2 dex difference between \Zdot\ and 1\% \Zdot\ models. The central $^{12}$C quickly drops as a result of the CN-cycle, before slightly increasing again during the CNO-cycle and remains constant before a slight upturn toward the end of the MS. The \Zdot\ model is most efficient at converting C to N, having the lowest relative abundance for the majority of the MS, but models of decreasing $Z$ have higher C due to their higher central temperatures which enable more rapid CNO-burning. Similarly, the central $^{16}$O abundance drops at the expense of N at the onset of the MS, with the 1\% \Zdot\ model having the lowest central O abundance during the MS evolution. This confirms that the lower $Z$ models have more efficient reactions during the MS due to their higher central temperatures. While the general behaviour of the more abundant CNO isotopes is similar for all Z, this is not always the case for heavier isotopes such as $^{23}$Na and $^{27}$Al. We find that all isotopes up to $^{23}$Na have the same general behaviour, yet from $^{23}$Na - $^{24,25}$Mg - $^{26,27}$Al there are differences in the reactions, mainly between $Z > $10\% \Zdot\ and 1\% \Zdot\ models. For instance, the central $^{23}$Na abundance appears to increase steadily for the majority of the MS, but decreases at the end of the MS for \ZSMC\ and 10\% \Zdot\ models.  

Interestingly, we find that the central $^{23}$Na abundance drops dramatically in the 1\% \Zdot\ model, already in the first 10\% of core H-burning. Proton capture on $^{23}$Na proceeds through resonances at the higher core temperatures for 1\% \Zdot\ (T$_{\rm{c}}$ $>$ 55MK, see Fig. \ref{fig:Tc_AllZ}), rapidly destroying the $^{23}$Na \citep{Boeltzig}. Moreover, the ratio between the $^{23}$Na(p,$\alpha$)$^{20}$Ne and the $^{23}$Na(p,$\gamma$)$^{24}$Mg reaction changes from over a factor 100 to closer to a factor 10 at these temperatures enhancing the feeding to Mg and Al. Similarly, the central $^{27}$Al abundance increases slightly during the MS for $Z > $1\% \Zdot\ models, but at 1\% \Zdot\ the behaviour is different with a dip mid-way through core H-burning before steeply increasing until core H-exhaustion. 

The reason for these differing behaviours is a combination of the production of each isotope relative to the initial $Z$ composition, which includes the abundance of the precursor isotopes in the Ne-Na and Mg-Al chains. The central temperatures are also slightly different at each $Z$, leading to changes in the efficiency of nuclear reactions. This means that the overall stellar properties may also play a role in the reaction rates and production of specific isotopes and net wind yields at lower $Z$. In general, the nuclear energy generation is more efficient at lower $Z$ due to the increased central temperature and density, which has the largest effect on reactions that are heavily T-dependent. The luminosity generated by the CNO cycle, log$_{\rm{10}}$ L$_{\rm{CNO}}$, attributed to the H-burning energy generation is increased from 6.2\Ldot\ at \Zdot\ to 6.3 at 1\% \Zdot, though the largest difference in L$_{\rm{CNO}}$ occurs in the remaining 20\% of H-burning where we see a 0.3dex difference at \Zdot, which correlates with the steep increase in $^{27}$Al from $Y_{c}$ = 0.8-1.0.
\begin{figure}
    \centering
    \includegraphics[width = \columnwidth]{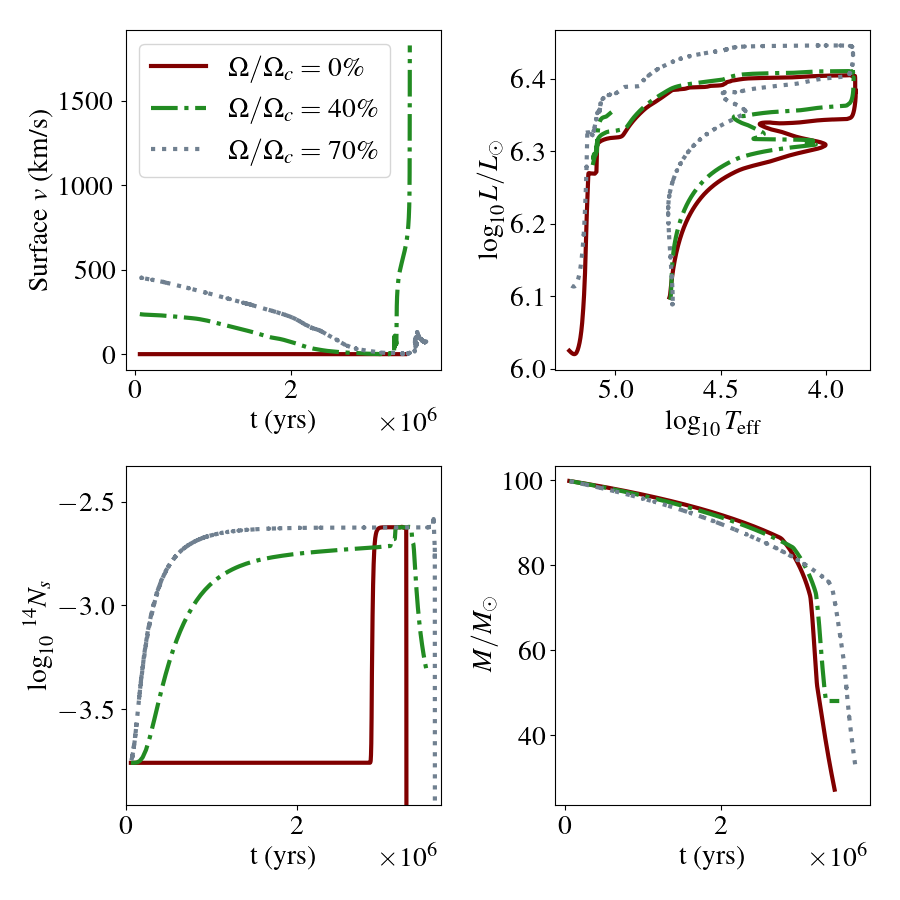}
    \caption{Surface rotation (top left), HRD (top right), surface $^{14}$N evolution (bottom left) and mass evolution (bottom right) of 100\Mdot\ models at \ZSMC\ for a range of critical rotation rates (0\% in solid red lines, 40\% in dash-dotted green lines and 70\% in dotted grey lines).}
    \label{fig:100M_SMC_panel}
\end{figure}

 While there are some changes to the nucleosynthesis of heavier elements at the lowest Z (1\% \Zdot), the Z-dependence on winds may be the dominant effect on yields from massive stars. For instance, while the central abundances may change by 0.3 dex across Z (\Zdot\ - 1\% \Zdot) for the more abundant CNO isotopes, and up to 1 dex for the less abundant Ne-Na and Mg-Al chains, the mass-loss rates are 1 dex lower at 1\% \Zdot\ compared to \Zdot\ during the first half of the MS, and up to 2 dex lower than \Zdot\ models during the second half of the MS (see Fig \ref{fig:mdot_MS_allZ}). Moreover, even if the central abundances changed significantly during the evolution of a low Z (1\% \Zdot) star, if it is not lost in the winds before becoming reprocessed into another isotope then it will not affect the yields (see Fig \ref{fig:yield100Z} bottom panel for example).

 \begin{figure}
    \centering
    \includegraphics[width = \columnwidth]{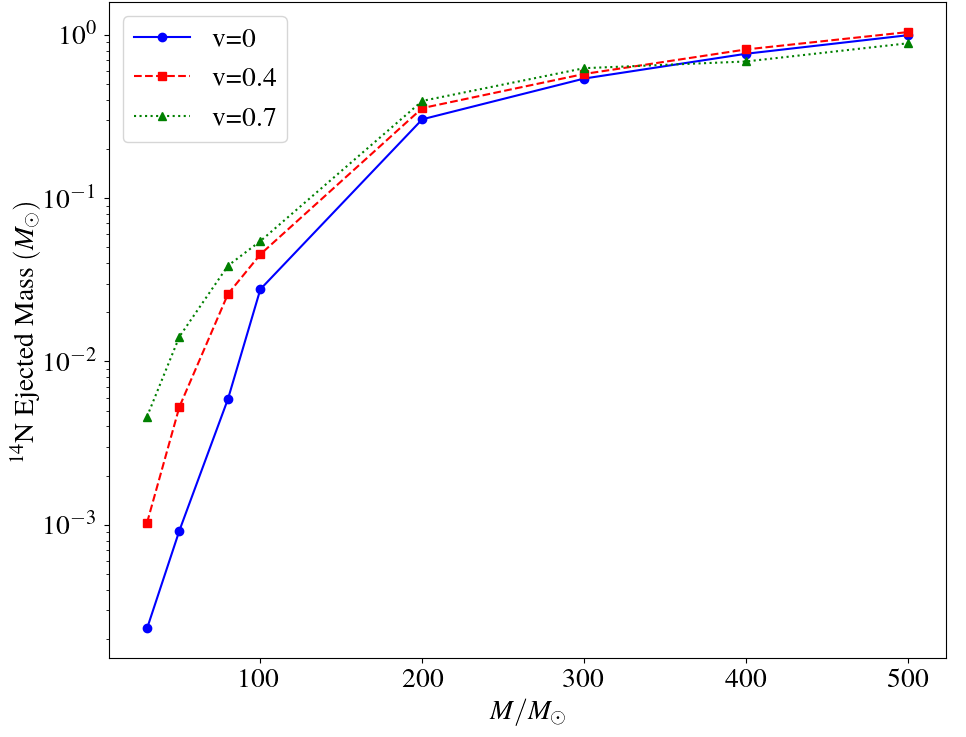}
    \caption{Ejected mass of $^{14}$N as a function of initial mass, both presented in solar masses, calculated at \ZSMC\ during the core H-burning phase. Each coloured line illustrates a different rotation rate where non-rotating models are shown in blue, 40\% critically rotating in red and 70\% critically rotating models in green.}
    \label{fig:N_MS}
\end{figure}
 \begin{figure}
    \centering
    \includegraphics[width = \columnwidth]{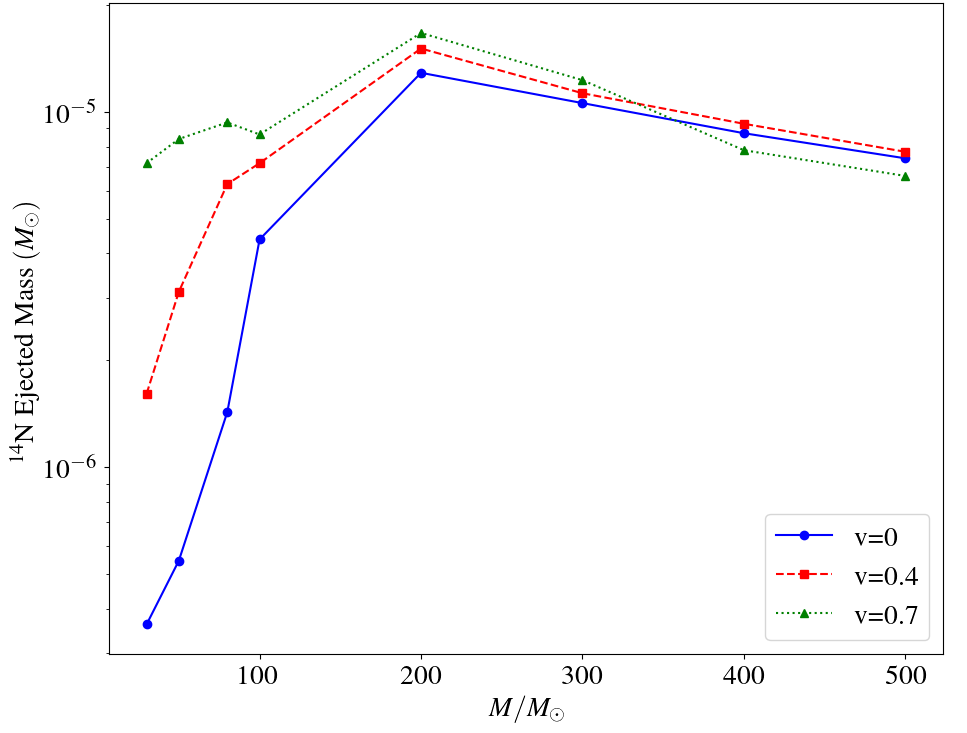}
    \caption{IMF-weighted ejected masses of $^{14}$N as a function of initial mass, both presented in solar masses, calculated at \ZSMC\ for the core H-burning phase. Each coloured line illustrates a different rotation rate where non-rotating models are shown in blue, 40\% critically rotating in red and 70\% critically rotating models in green.}
    \label{fig:N__MS_IMF}
\end{figure}
\section{Rotation effects at low Z}
In this section we present the wind yields of our \ZSMC\ and 10\% \Zdot\ models which are calculated for 3 rotation rates (non-rotating, 40\% critical and 70\% critical). As discussed in Sect. \ref{masslossrotation}, we implement the increase in mass-loss rates due to the proximity to the $\Omega\Gamma$-limit as described by \cite{MM00}.

\begin{figure}
    \centering
    \includegraphics[width = 1.1\columnwidth]{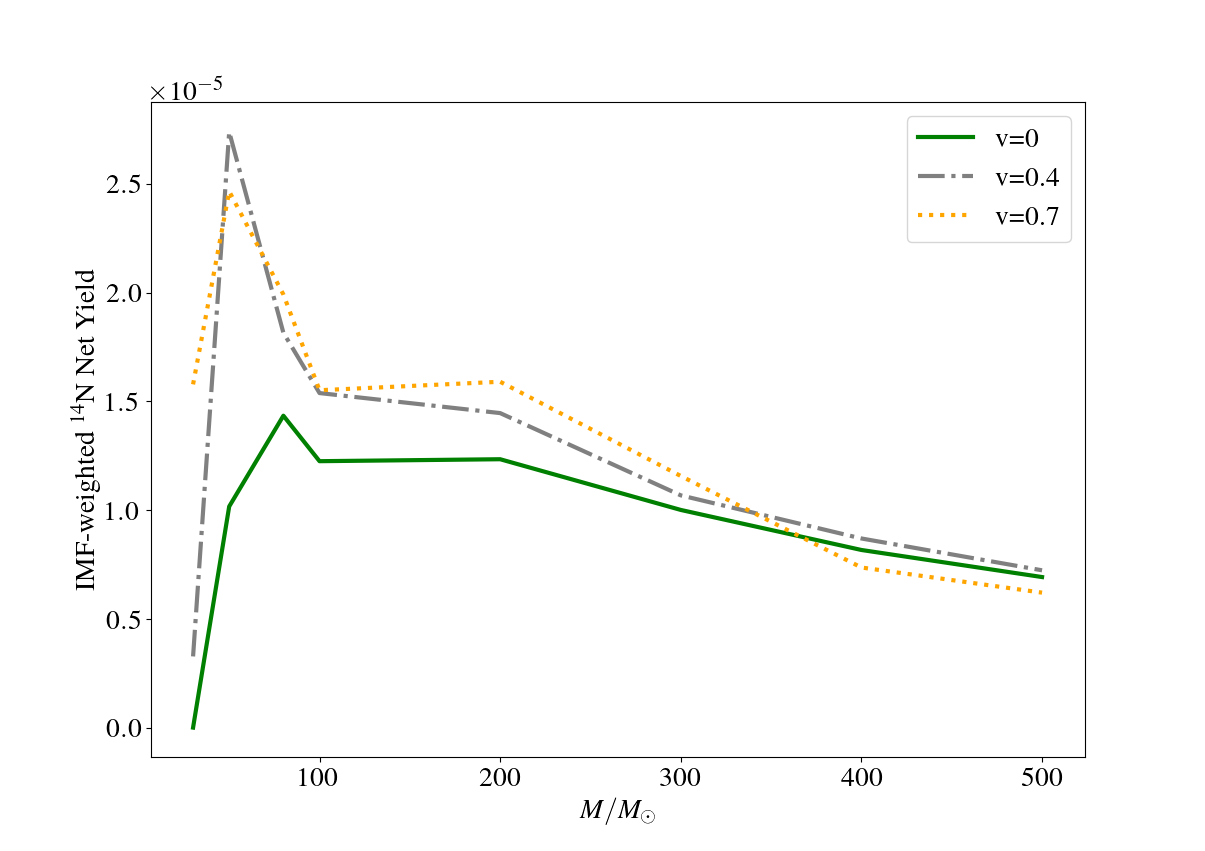}
    \caption{IMF-weighted net wind yields of $^{14}$N as a function of initial mass, both presented in solar masses, calculated for the H and He-burning phases, at \ZSMC. Each coloured line illustrates a different rotation rate where non-rotating models are shown in green, 40\% critically rotating in grey and 70\% critically rotating models in orange.}
    \label{fig:N_IMF}
\end{figure}

\begin{figure*}
    \centering
    \includegraphics[width = \textwidth]{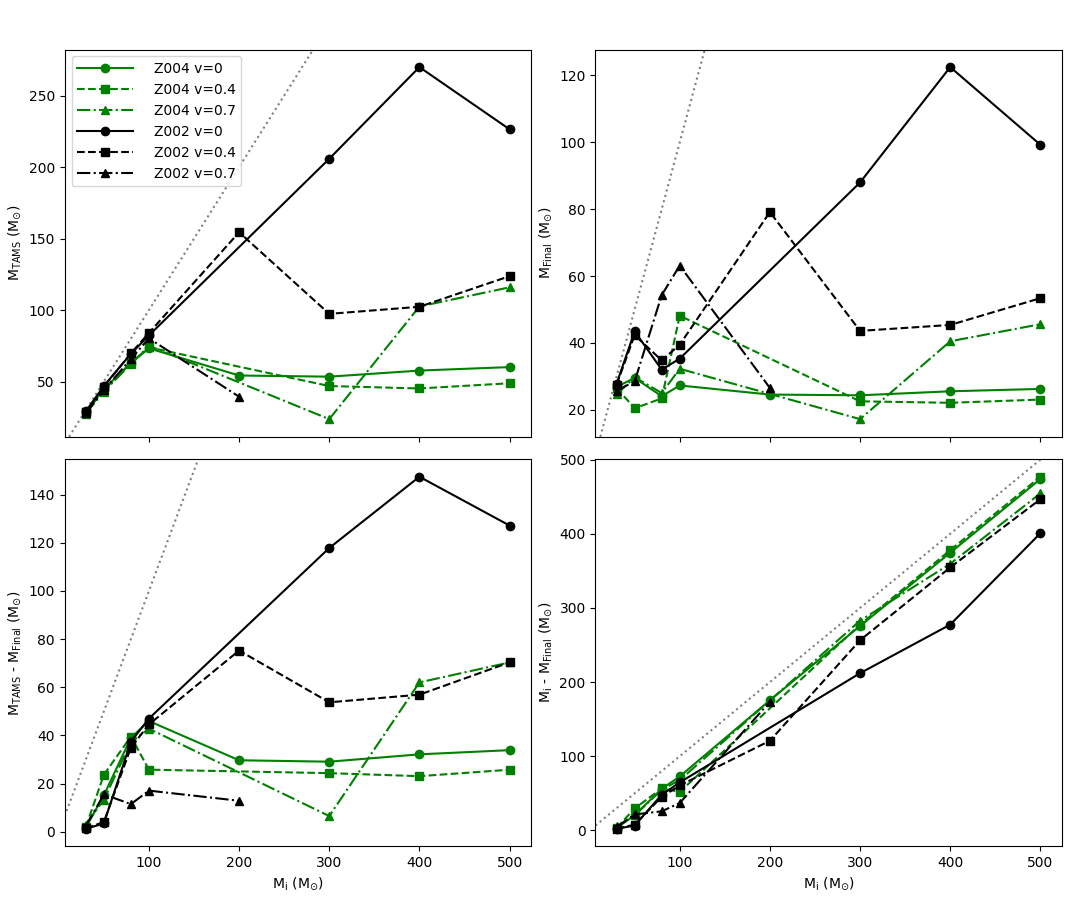}
    \caption{Stellar mass properties at the end of core H-burning ($M_{\rm{TAMS}}$), and end of core He-burning (final) for models calculated at \ZSMC\ (green) and 10\% \Zdot\ (black) for a range of critical rotation rates (0\% in solid lines, 40\% in dashed lines and 70\% in dash-dotted lines). A 1:1 mass ratio has been plotted for each panel in grey dotted lines. }
    \label{fig:zsmc01z_prop}
\end{figure*}

\subsection{\ZSMC}
At 20\% of \Zdot, the SMC is an excellent nearby laboratory for testing the evolution of massive stars at low Z. While the effects of stellar winds are reduced, VMS (M $>$ 100\Mdot) still lose 20-80\% of their total mass on the MS and massive stars (30\Mdot\ $<$ M $<$ 100\Mdot) lose 20-60\% of their mass in the post-MS, leaving $\sim$ 10-40\% of their initial mass. Table \ref{tab:SMC_yields} provides the full list of ejected masses and net yields for isotopes of $^{1}$H to $^{28}$Si, from the ZAMS until core He-exhaustion, for 3 sets of critical rotation rates (0 top, 40\% middle, and 70\% bottom). We find that the amount of C, N, O, Ne, Na, and Mg ejected are still significant for all rotation rates (10$^{0}$ to 10$^{-2}$ \Mdot).
We explore 100\Mdot\ models at \ZSMC\ for a range of rotation rates in Fig. \ref{fig:100M_SMC_panel}. The HRD of these models (upper right) show that all 100\Mdot\ models evolve bluewards after core H-exhaustion at log T$_{\rm{eff}}$ $\sim$ 4.0 and continue towards stripped Helium stars forming cWRs. Interestingly, the mass-loss rates of the 40\% critically rotating model are not sufficiently high enough to prevent spin up at the contraction phase, while the 70\% critically rotating model loses enough mass and angular momentum to spin down throughout its evolution (upper left in Fig. \ref{fig:100M_SMC_panel}). The surface enrichment of $^{14}$N increases with rotation, partly due to increased mass loss yet also due to increased rotational mixing dredging more $^{14}$N to the surface early on the MS. This illustrates that with increased rotation, stars will eject significantly more $^{14}$N throughout their MS evolution.

Figure \ref{fig:N_MS} shows the ejected mass of $^{14}$N from 30-500\Mdot\ stars over the core H-burning phase, for each rotation rate. Regardless of initial rotation rate (including non-rotating), the 200-500\Mdot\ stars can contribute a significant amount of $^{14}$N in the range 0.1-1\Mdot\ per individual VMS \citep{Vink23}. Below the transition point (M$<$100\Mdot), where stars drive optically thin winds rather than the optically thick winds experienced by VMS, the ejecta of $^{14}$N quickly drops. However, with rapid rotation, these 50-100\Mdot\ stars can still eject a relevant quantity of $^{14}$N. When weighted by an IMF (M$^{-1.9}$) the contribution from 200\Mdot\ stars remains dominant (Fig.\ref{fig:N__MS_IMF}). Moreover, canonical O stars with masses 30-100\Mdot\ require extreme rotation rates of $\Omega_{\rm{ini}} / \Omega_{\rm{crit}} = 0.7$ to compare with the ejecta of 300-500\Mdot\ stars.

When the total core H and He-burning phases are considered, it is the rotating stars just below the transition point (50-80\Mdot) which dominate the IMF-weighted ejecta (Fig.\ref{fig:N_IMF}). The total contribution from this mass range outweighs that of stars beyond 100\Mdot, as well as canonical O stars below 50\Mdot. Moreover, we show the IMF-weighted yields of $^{14}$N in Fig. \ref{fig:N_IMF} highlighting a peak in $^{14}$N ejecta from stars with initial masses 80-100\Mdot. The maximum $^{14}$N IMF-weighted net yields come from this mass range, for all rotation rates, but particularly with rotation rates of 40\% critical, while in the mass range of 100-300\Mdot\ the 70\% critically rotating models eject the most $^{14}$N for these masses.

In general, rotating massive stars lose more mass than non-rotating stars due to the proximity of the $\Omega\Gamma$-limit which increases mass-loss rates due to the von Zeipel effect. Interestingly, at 70\% critical rotation we find that VMS eject substantial amounts of C and O, $\sim$ 10\Mdot\ more compared to the non-rotating and 40\% critically rotating models. The non-rotating and 70\% critically rotating models eject less $^{14}$N than the 40\% rotating models since the rotational mixing effects lead to higher $^{14}$N enrichment at the surface earlier in the evolution. But the highest rotation rates also lead to higher mass-loss rates on the post-MS ejecting more C and O. At the highest masses ($\sim$ 200-500\Mdot) the rotation rate does not appear to have an effect on the yields of Ne, Na, Mg, Al or Si. However, at the lower mass range $\sim$30-80\Mdot\ rotating stars provide an order of magnitude more $^{14}$N than non-rotating stars when IMF-weighted over the MS phase (Fig. \ref{fig:N__MS_IMF}).

\begin{figure}
    \centering
    \includegraphics[width = \columnwidth]{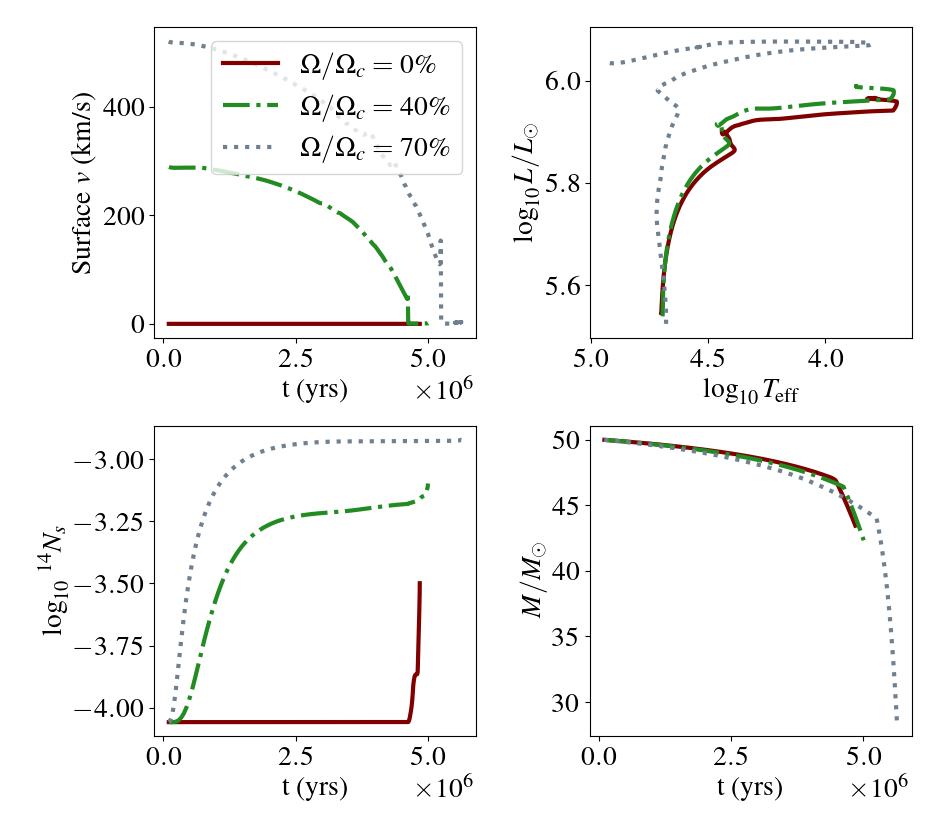}
    \caption{Surface rotation (top left), HRD (top right), surface $^{14}$N evolution (bottom left) and mass evolution (bottom right) of 50\Mdot\ models at 10\% \Zdot\ for a range of critical rotation rates (0\% in solid red lines, 40\% in dash-dotted green lines and 70\% in dotted grey lines).}
    \label{fig:50M_01z_panel}
\end{figure}

\begin{figure}
    \centering
    \includegraphics[width = \columnwidth]{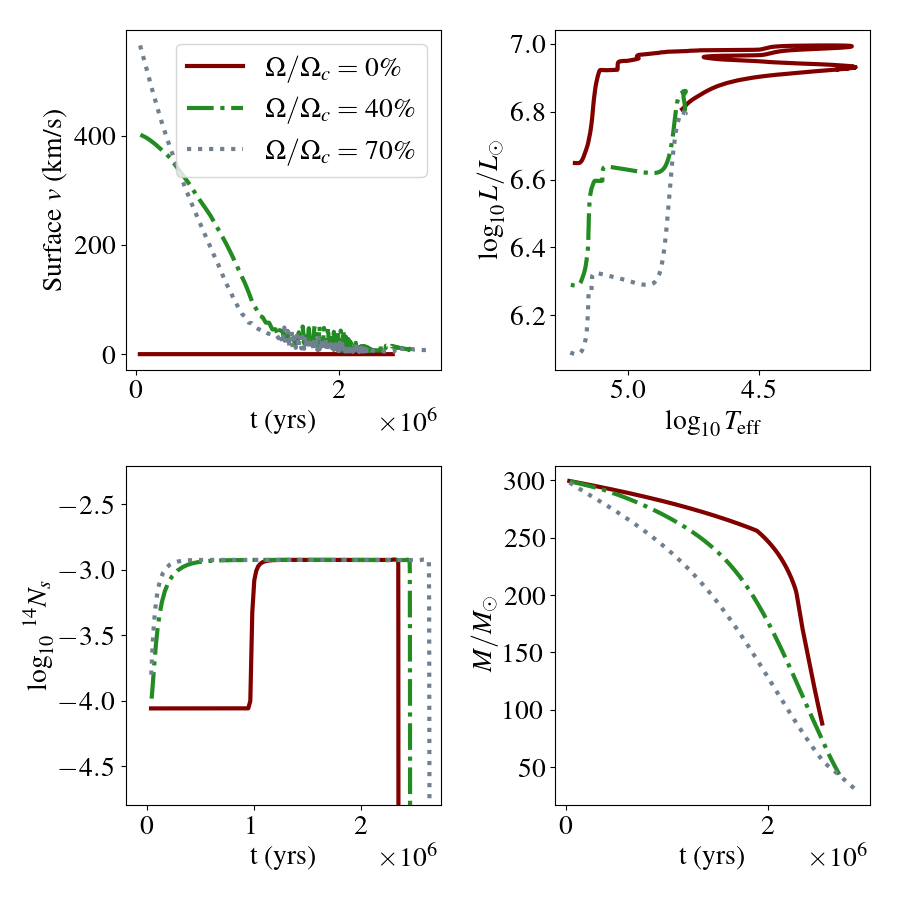}
    \caption{Surface rotation (top left), HRD (top right), surface $^{14}$N evolution (bottom left) and mass evolution (bottom right) of 300\Mdot\ models at 10\% \Zdot\ for a range of critical rotation rates (0\% in solid red lines, 40\% in dash-dotted green lines and 70\% in dotted grey lines).}
    \label{fig:300M_01z_panel}
\end{figure}

\subsection{1/10th \Zdot}

Figure \ref{fig:zsmc01z_prop} shows the stellar mass properties at both the end of core H-burning and the end of core He-burning as a function of initial stellar mass for models with two different metallicities:  \ZSMC\ (green lines) and 10\% \Zdot\  (black lines). The different line styles represent variations in the rotation rate, where solid lines correspond to zero rotation, dashed lines represent 40\% of the critical rotation, and dash-dotted lines indicate 70\% of the critical velocity. Across all panels, the most noticeable effect is that mass loss increases significantly with higher initial mass and metallicity. Stars at 10\% \Zdot\ lose less mass compared to those with \ZSMC\, especially for stars with initial masses beyond $\sim$100 \Mdot. Furthermore, the rotation rate significantly influences mass loss, with higher rotation rates (dash-dotted lines) leading to much lower final masses, especially at masses M $<$ 300\Mdot. The indirect effects of rotation become important for stars with initial masses above $\sim$ 100 \Mdot, where models with 70\% of the critical rotation rate lose more mass due to the $\Omega\Gamma$-limit early in the evolution. Subsequently stripping the star of its angular momentum, the resulting mass at core H-exhaustion or final masses are much higher than stars which began their evolution with 40\% critical rotation. The bottom left panel, which shows the same behaviour as the upper 2 panels demonstrate that rotation plays a role throughout the evolution by directly increasing the mass-loss rate, but also indirectly by increasing the luminosity and temperature of rotating models leading to increased mass loss even on the post-MS where most of the angular momentum has already been lost. By comparing the masses at core H-exhaustion with final masses (top left and right panels) we see that stars still lose a significant amount of mass on the post-MS (see the gradient of the 1:1 mass ratio shown by grey dotted lines). The bottom right panel indicates that despite rotationally-induced mass loss, the total mass lost as a function of initial mass remains relatively linear, though the 10\% \Zdot\ models lose up to 100\Mdot\ less than \ZSMC\ models at the highest mass range ($\sim$ 400-500\Mdot).

Figures \ref{fig:50M_01z_panel} and \ref{fig:300M_01z_panel} show the effects of rotation on the evolution of 50 $M_\odot$ and 300 $M_\odot$ stars at 10\% \Zdot. The 50 $M_\odot$ models (Fig. \ref{fig:50M_01z_panel}) follow a standard O-star mass loss prescription, while the 300 $M_\odot$ models (Fig. \ref{fig:300M_01z_panel}) are subject to enhanced optically thick winds, resulting in notable differences in mass loss and rotational evolution.

In the 50 $M_\odot$ models, rotation plays a moderate role in altering surface properties. As shown in the upper left panel, higher rotation rates ($ \Omega_{\rm{ini}} / \Omega_{\text{crit}} = 40 \%$ and $ 70 \%$) maintain relatively high surface velocities for the entire H-burning phase. By the end of core He-burning, even the most rapidly rotating models exhibit significant angular momentum loss as a result of wind mass loss. The HRD (upper right panel) reveals that the faster rotating models shift to slightly higher luminosities and hotter effective temperatures as a result of rotational mixing. However, the O-star wind rates limit the overall loss of angular momentum, leading to significant surface $^{14}$N enrichment, and relatively minor differences in mass evolution (for non-rotating and 40\% critical models). The 70\% critical rotation model experiences significant chemical mixing and evolves bluewards to hotter temperatures towards the end of core He-burning, and loses a substantial amount of mass in this burning phase.

In contrast, the 300 $M_\odot$ models display a much more significant impact of rotationally-induced mass loss since these stars are above the transition point and already experience enhanced optically thick winds with high mass-loss rates. Hence, the combined effect of rotation and mass loss can play a key role in the evolution of stars near the transition point at low metallicity. As seen in the surface velocity evolution (upper left panel), stars with higher rotation rates experience a rapid decline in surface velocity from the onset of core H-burning, driven by the strong mass loss from optically thick winds. This results in the 70\% rotating model losing a considerable portion of its angular momentum early on, leading to a rapid decrease in surface rotation. The HRD (upper right panel) shows that rotating models evolve closer to the transition point causing the switch from horizontal evolution (non-rotating) to the vertical evolution of VMS seen at predominantly higher Z, with a dramatic drop in luminosity and subsequent WR evolution. The mass evolution plot (bottom right panel) highlights the impact of rotationally-enhanced winds, with the 70\% rotating model experiencing significant mass loss during the MS, leading to a 50-100\Mdot\ difference with the non-rotating model during the MS. Additionally, the surface $^{14}$N enrichment is more pronounced in the 300 $M_\odot$ model (bottom left panel), with rapidly rotating models showing a very quick increase by a factor of 10 in $^{14}$N abundance. This rapid increase in $^{14}$N is due to higher central temperatures (than lower mass or higher Z stars) reaching CN-equilibrium almost instantly, stripping of the outer pristine layers quickly, and the effects of rotational mixing dredging $^{14}$N to the surface. Comparable figures \ref{fig:50M_SMC_panel} and \ref{fig:300M_SMC_panel} for 50\Mdot\ and 300\Mdot\ models at \ZSMC\ are shown in Appendix \ref{figappendix}. These figures illustrate that at higher Z, the effects of rotation are diminished slightly given the increased mass loss, particularly for the 300\Mdot\ models which are now far beyond the transition point.

\begin{figure}
    \centering
    \includegraphics[width = \columnwidth]{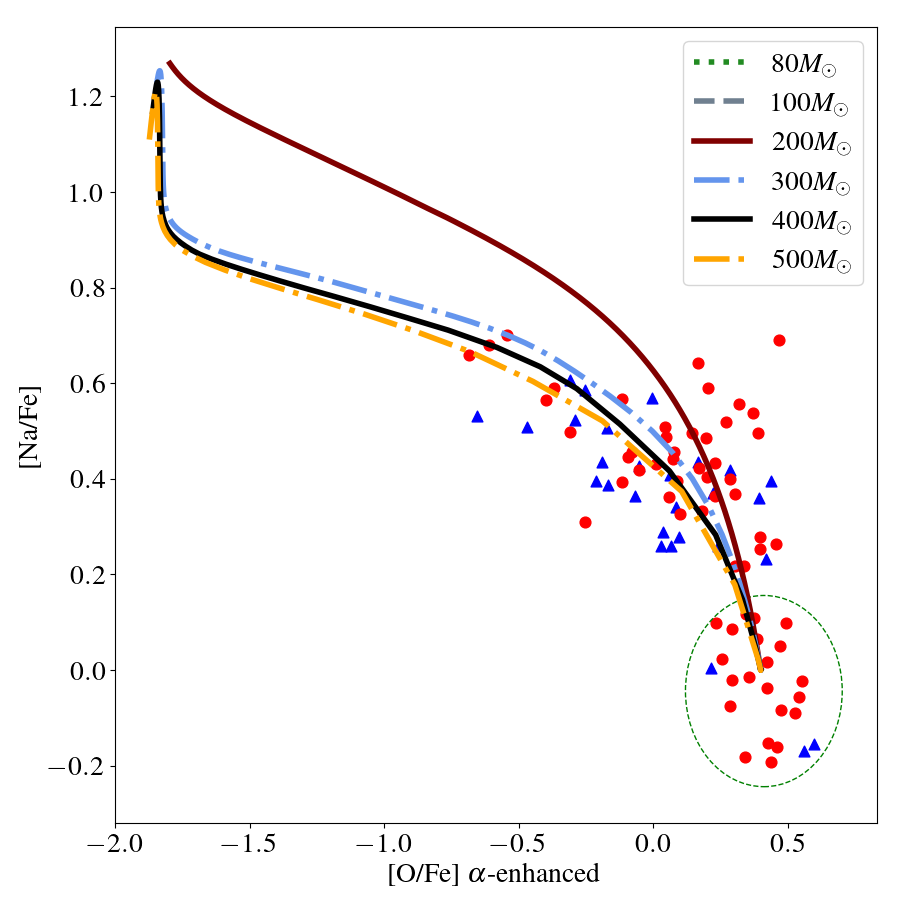}
    \caption{[Na/Fe] and [O/Fe] surface evolution for 80-500\Mdot\ models during core H-burning, at 1/30th \Zdot. The red circles represent detections of stars in NGC 5904 while blue triangles represent upper limits in [O/Fe] as defined by \citet{carretta09}. The green dashed circle establishes the abundance ratios of field stars ([O/Na] $=$ 0.6), with stars outside this representing P2 stars.}
    \label{fig:MS_NaO}
\end{figure}

\begin{figure}
    \centering
    \includegraphics[width = \columnwidth]{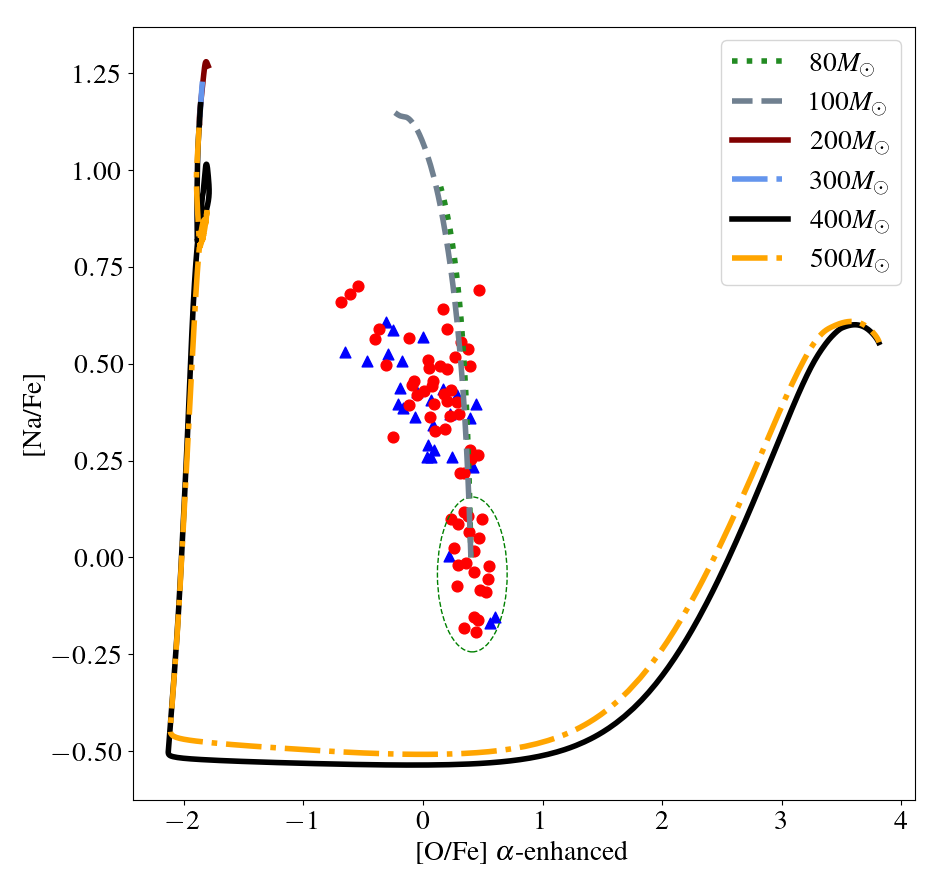}
    \caption{[Na/Fe] as a function of [O/Fe] for 80-500\Mdot\ models during core He-burning, at 1/30th \Zdot. First and second population stars from the cluster NGC5904, adopted from \citet{carretta09} are shown in red circles and blue triangles respectively.}
    \label{fig:postMS_NaO}
\end{figure}
\section{Globular clusters}

In this section we address the importance of globular clusters at $Z \sim 3\%$ \Zdot\ for stellar nucleosynthesis and evolution. Such low metallicity GCs are common in the outer regions of galaxies, including our Milky Way, and are considered to be remnants from the early universe, providing a glimpse into the conditions of early star formation. One of the most significant chemical peculiarities observed in GCs is the Na-O anti-correlation. This refers to the inverse relationship between the abundance of sodium (Na) and oxygen (O) in stars within the same cluster. Stars with high Na content typically have low O content, and vice versa. This pattern is not seen in field stars, making it unique to GCs, and is expected to be caused by multiple stellar populations. These populations are characterised by different chemical compositions, including the Na-O anti-correlation. Population 1 (P1) stars generally show abundances similar to field stars of similar metallicity, while Population 2 (P2) stars exhibit enhanced Na and depleted O, suggesting a different chemical enrichment history. The distinction between P1 and P2 stars suggests a complex formation history for GCs. It implies that clusters formed multiple populations of stars over time, with the second population forming from material that was chemically enriched.

\cite{carretta09} explore the Na-O anti-correlation in 15 GCs and the distinction between different stellar populations within these clusters. They define P1 stars as those with similar [Na/Fe] and [O/Fe] ratios as field stars, while stars with enhanced Na/Fe and depleted O/Fe are considered P2 stars. We investigate the relevance of our 80-500\Mdot\ models in forming P2 stars by exploring the surface evolution of [Na/Fe] and [O/Fe] in relation to the observations of stars in NGC 5904. While this cluster is not atypical for GCs, it provides a broad range of [Na/Fe] and [O/Fe] to test our evolutionary models against. Moreover, the UVES observed [Fe/H] abundance of NGC 5904 (-1.34) is close to our model grid (-1.35). We note that since our initial composition does not account for the $\alpha$-enhancement observed \citep[e.g.][]{Gratton12}, we have corrected our model data in [O/Fe] by increasing the ratio at each point by $+$0.4 dex. Figure \ref{fig:MS_NaO} illustrates the MS evolution of the surface composition in our models with respect to the abundance ratios of observed stars in NGC 5904. We find that stars with M$_{i}$ $\geq$ 200\Mdot\ qualitatively explain the P2 stars lying outside the green circle which shows the abundances of field stars (P1) where [O/Na] $\simeq$ 0.6 \citep{charbonnel16}. As the models evolve during core H-burning, their surfaces enrich further in Na and deplete in O. All models begin with the same abundance ratios relative to solar (0, 0). The higher mass models become more O-depleted than lower mass models, and also display increased Na-enrichment. 

Similarly, we investigate the surface evolution of Na-O during the post-MS in Fig. \ref{fig:postMS_NaO} finding that stars with M$_{i}$ $\sim$ 80-100\Mdot\ evolve with surface Na and O ratios that encompass the observed Na/O pattern. These figures illustrate the effects of winds on the surface evolution and ejecta of M $>$ 80\Mdot\ stars. As a result of stronger mass loss on the MS, higher mass models (M $>$ 200\Mdot) lie beyond the observed [Na/O] pattern, with extremely O-depleted surfaces until both O and Na enrich as a result of He-burning nucleosynthesis. However, even when these VMS models enrich in O during core He-burning, the relative [O/Fe] abundances are orders of magnitude higher than observed P2 stars. Therefore, the best representation of observed NGC 5904 abundances appears to be VMS with M $\geq$ 200\Mdot\ on the MS, or 80-100\Mdot\ stars on the post-MS. At low Z (3\% \Zdot), MS winds are reduced for stars below the transition point (M$\leq$200\Mdot) and therefore do not showcase the H-processed Na and O until much later in their evolution. On the other hand, VMS (M$\geq$200\Mdot) strip their outer layers more quickly to reveal the Na-rich and O-depleted material early on the MS.

\begin{table}
    \caption{Net wind yields of $^{16}$O and $^{23}$Na for models calculated at 3$\% $\Zdot\ and initial masses of 30-500\Mdot\ in solar masses, calculated from the onset of core H-burning until core H-exhaustion.}
    \centering
    \begin{tabular}{ccc}
    \hline
     $M_{\rm{i}}/\rm{M}_{\odot}$	&$^{16}$O&	$^{23}$Na\\
\hline \hline
        30 & 2.03E-18 & 7.18E-21 \\
        50 & 1.31E-17 & 2.53E-20 \\
        80 & -8.86E-14 & 3.29E-16  \\
        100 & -1.52E-13 & 8.98E-14  \\
        200 & -4.07E-03 & 2.41E-04  \\
        300 & -1.27E-02 & 8.15E-04  \\
        400 & -2.01E-02 & 1.23E-03  \\
        500 & -2.87E-02 & 1.66E-03  \\

\hline
    \end{tabular}
    \label{tab:GC_EMs}
\end{table}

Figures \ref{fig:MS_NaO} and \ref{fig:postMS_NaO} illustrate the evolution of surface abundances of our models during the MS and post-MS, which provides a comparison to the observed abundance ratios in GCs like NGC 5904. While these surface properties highlight the relevance of VMS in reproducing the abundance patterns of Na and O in GCs, they do not directly correlate to the ejecta of these stars which could form a second population of stars. Therefore, we also present the net wind yields of $^{16}$O and $^{23}$Na, relative to their initial composition, for each of our models integrated over the H-burning phase. Table \ref{tab:GC_EMs} confirms that on the MS, stars with M$_{i}$ $\geq$ 100\Mdot\ eject O-depleted, Na-rich material with up to 10$^{-3}$ \Mdot\ of Na ejected for the highest mass models.

This scenario changes for VMS when the post-MS is included in the total net yields of Na and O, integrating the abundances of mass lost from the onset of core H-burning until core He-exhaustion.
% (see Table \ref{tab:appZ0006EM}). 
At this point, sufficient O-rich material is ejected (10$^{-4}$ \Mdot) to result in positive net yields of O. Therefore, we find that the MS yields of VMS best represent the observed trend in Na-O abundances. Crucially, at such low Z, the net wind yields of stars with M$<$ 100\Mdot\ are 10 orders of magnitude lower than stars above this mass range, suggesting that massive O stars cannot produce enough Na-rich and O-poor material compared to VMS, to produce a second population of stars with such abundance ratios.
\begin{figure}
    \centering
    \includegraphics[width = \columnwidth]{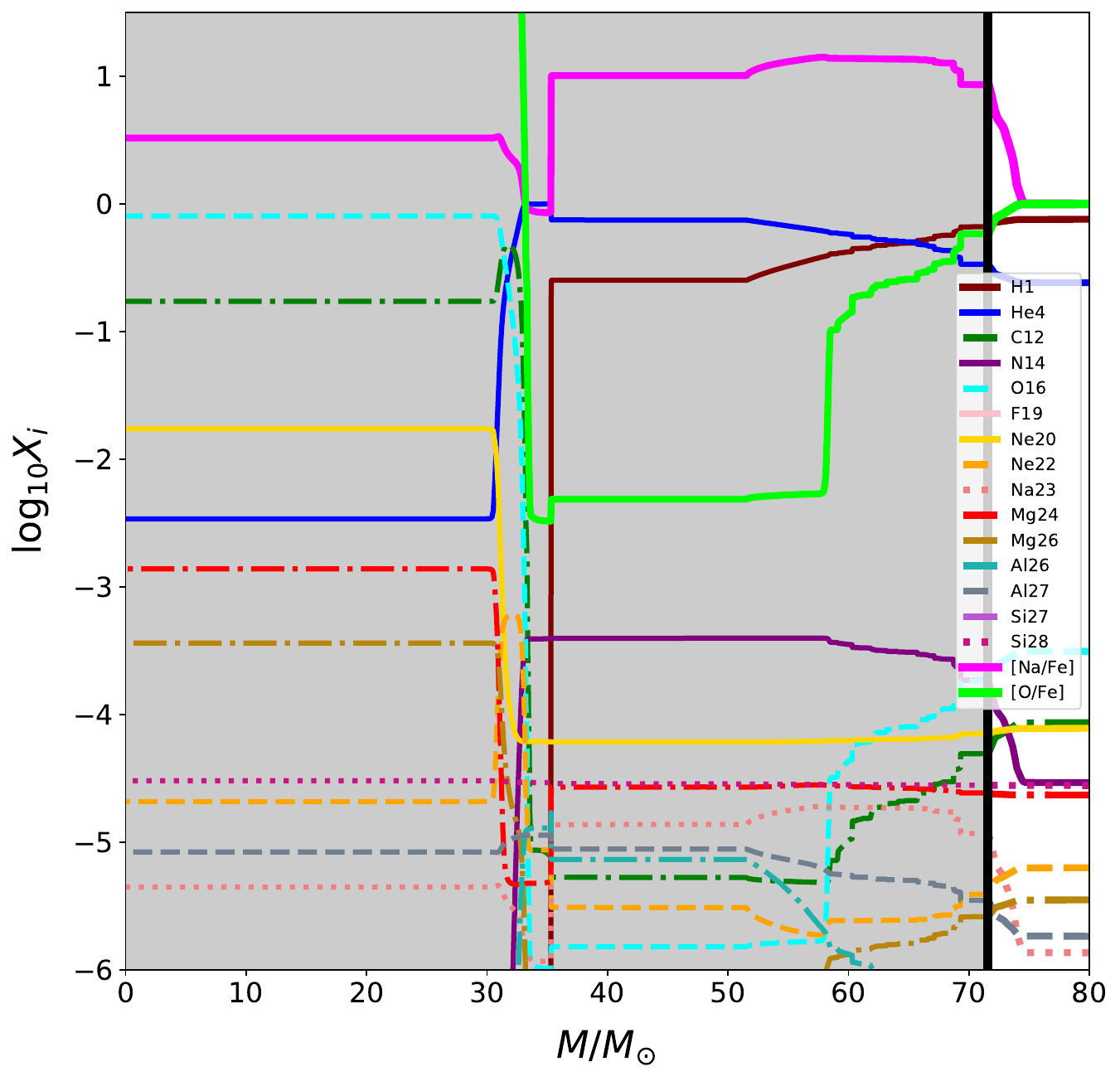}
    \caption{Time evolution of surface abundances of a 80\Mdot\ model at 3\% \Zdot, calculated from the onset of core H-burning until core He-exhaustion. [Na/Fe] and [O/Fe] are included.}
    \label{fig:80NaO}
\end{figure}
\begin{figure}
    \centering
    \includegraphics[width = \columnwidth]{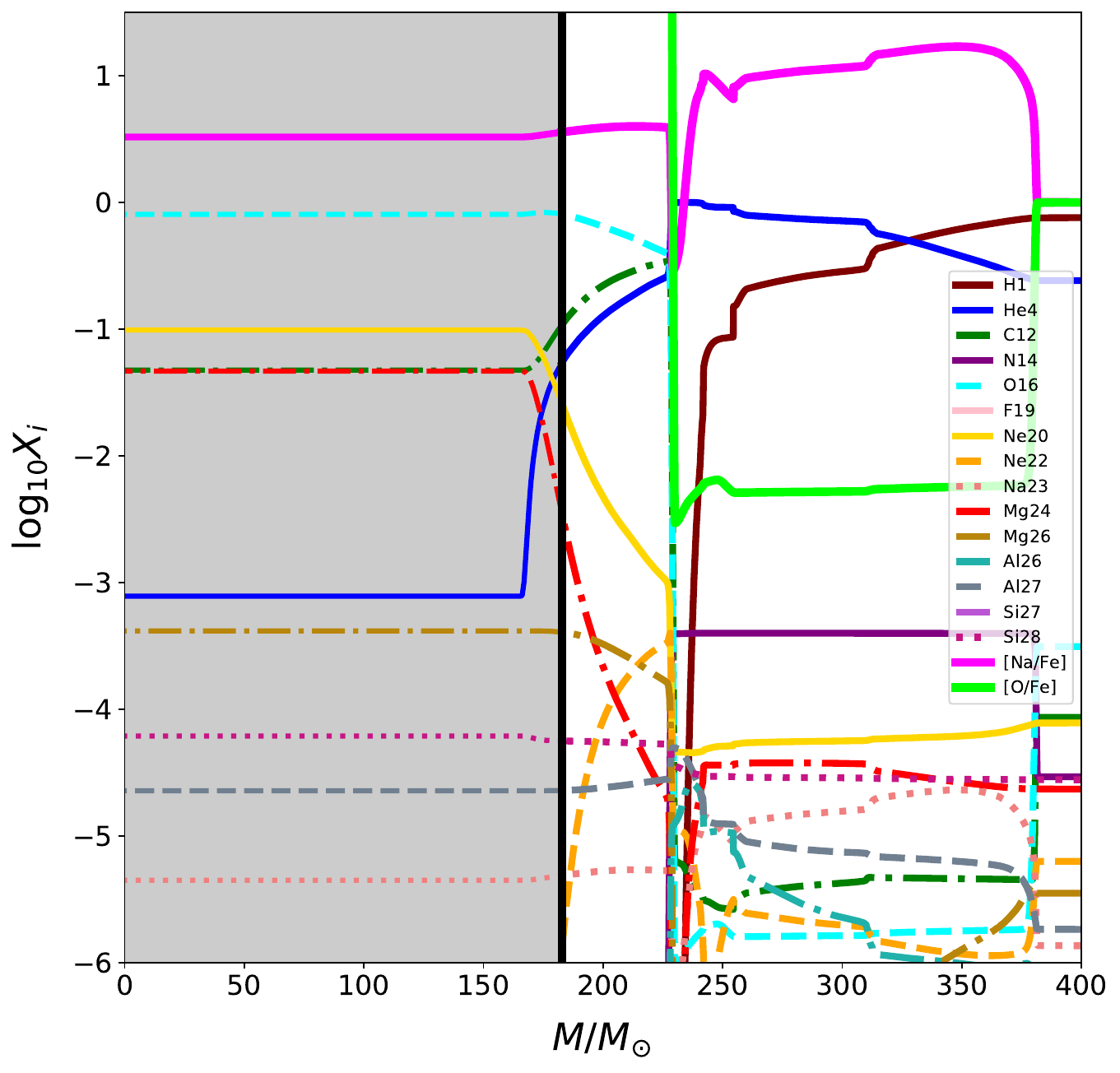}
    \caption{Time evolution of surface abundances of a 400\Mdot\ model at 3\% \Zdot, calculated from the onset of core H-burning until core He-exhaustion. [Na/Fe] and [O/Fe] are included.}
    \label{fig:400NaO}
\end{figure}
Another way to consider the contribution of [Na/Fe] enhanced and [O/Fe] depleted material from these stellar models is to investigate the interior and surface composition in mass fraction with time. Figures \ref{fig:80NaO} and \ref{fig:400NaO} show the same surface composition evolution as in Fig. \ref{fig:yield100Z}, where the interior composition remaining at the end of core He-burning is shown in the grey shaded region. Though we now also include the [Na/Fe] and [O/Fe] ratios in pink and green respectively, starting at 0 (right) and evolving (left) as mass is lost from the outer layers. We can see in Fig. \ref{fig:80NaO} that there is a small region ($\sim$72\Mdot) where the surface is [Na/Fe]-rich, but only after the black solid line, which represents the end of core He-burning. Within the grey shaded interior composition from surface to centre (right to left), there is a significant drop in [O/Fe] with a simultaneous rise in [Na/Fe] which then remains constant in the region 35-60\Mdot\ in the stellar interior. This means that during 99\% of the stellar lifetime, an 80\Mdot\ star does not eject a large amount of material (at this $Z$) which matches the observed Na-O pattern, but if the star did experience enhanced mass loss towards the end of its life through eruptions or another mechanism, then the interior composition which is representative of the Na-O pattern could be lost to the ISM. On the other hand, a 400\Mdot\ star with the same $Z$ and initial composition shown in Fig. \ref{fig:400NaO} produces a similar portion of [Na/Fe]-rich and [O/Fe]-poor material (see region $\sim$ 220-370\Mdot). However, the plateau in both abundance ratios occurs over a larger mass range ($\sim$ 150\Mdot\ compared to $\sim$ 30\Mdot in the 80\Mdot\ model) and crucially, is lost in the ejecta (white region) rather than being maintained in the stellar interior (grey region).

\begin{figure}
    \centering
    \includegraphics[width = \columnwidth]{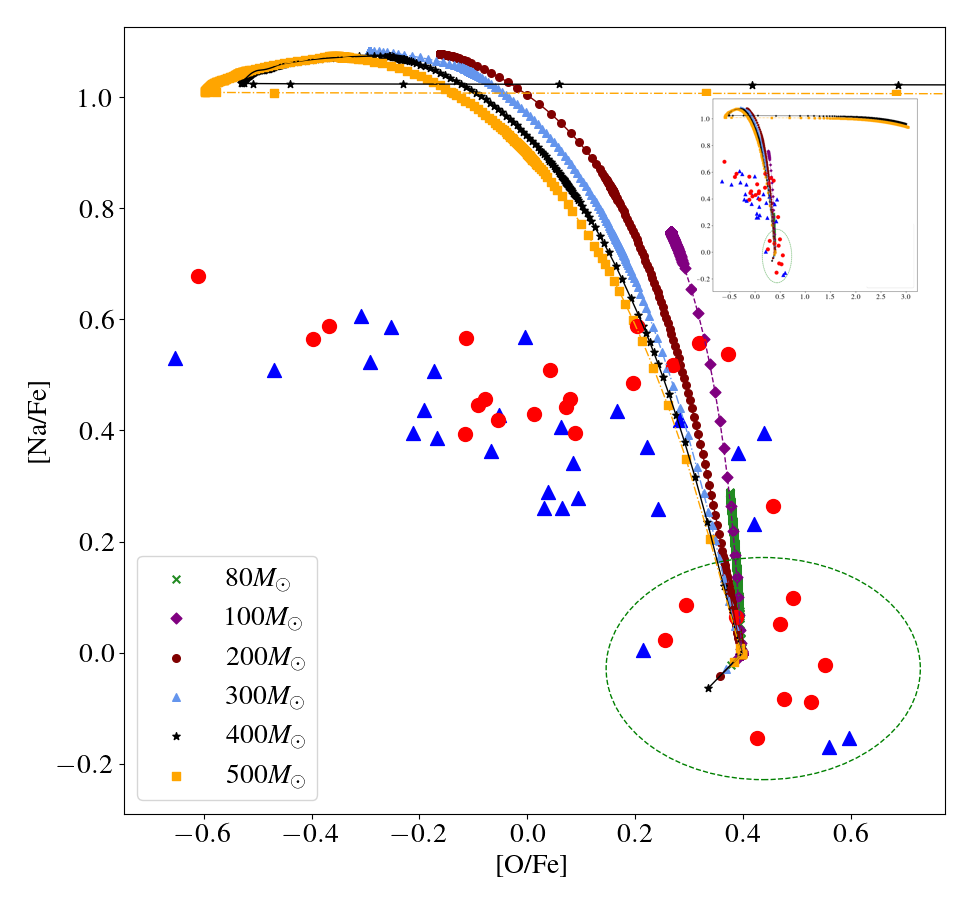}
    \caption{Cumulative ejected masses of [Na/Fe] and [O/Fe] for 80-500\Mdot\ models calculated at 1/30th \Zdot. Observations of stars in NGC 5904 are shown as in Fig. \ref {fig:MS_NaO}, adopted from \citet{carretta09}. The insert includes the wide range of O-enrichment ejected by the 400-500\Mdot\ models during the post-MS.}
    \label{fig:GC_EM_hist}
\end{figure}
 \begin{figure}
    \centering
    \includegraphics[width =\columnwidth]{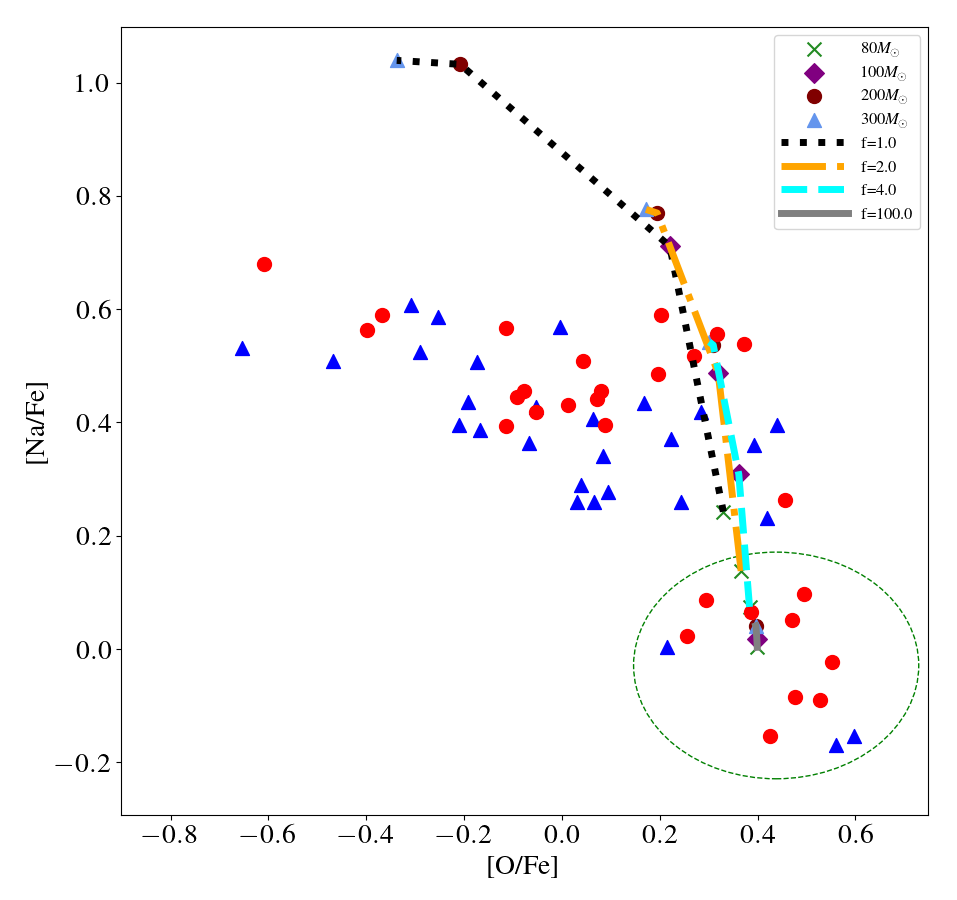}
    \caption{Diluted ejected from 80-500\Mdot\ models calculated at 1/30th \Zdot. Observations of stars in NGC 5904 are shown as in Fig. \ref {fig:MS_NaO}, adopted from \citet{carretta09}. Dilution factors ranging from 1 to 100 are shown for each model (assorted symbols) and connected for each f value (coloured lines).}
    \label{fig:dilution}
\end{figure}
 \begin{figure}
    \centering
    \includegraphics[width =\columnwidth]{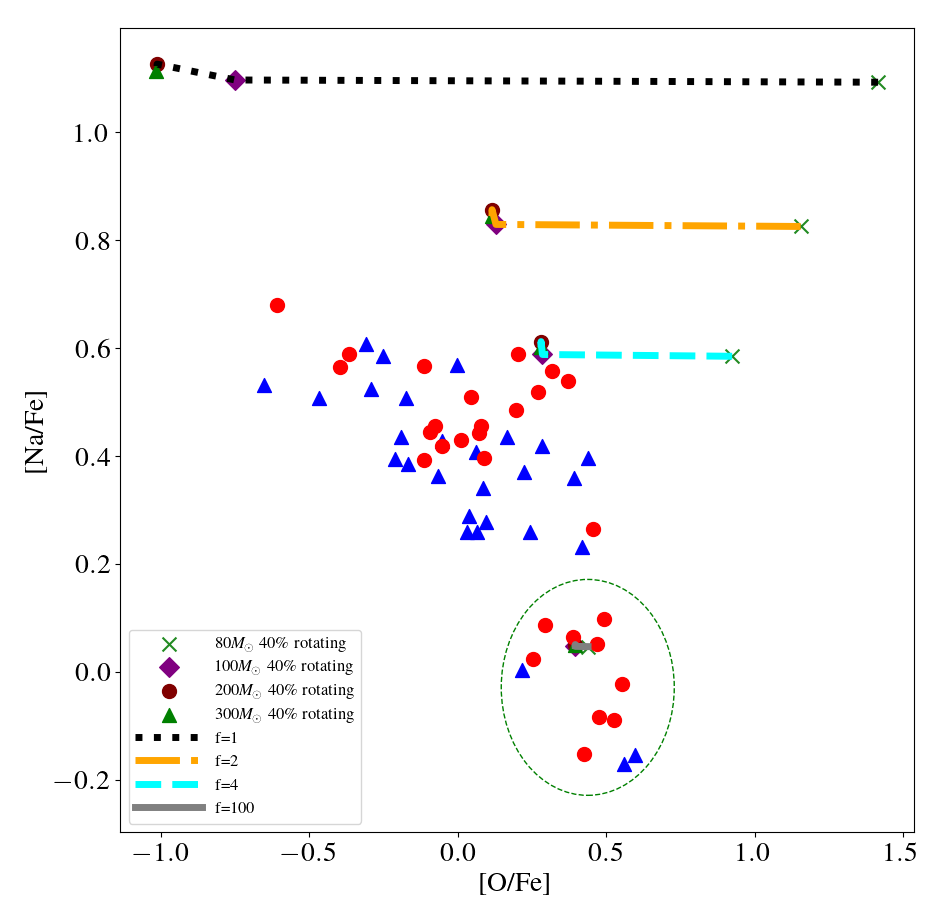}
    \caption{Diluted ejected from initially 40\% critically rotating 80-300\Mdot\ models calculated at 1/30th \Zdot. Observations of stars in NGC 5904 are shown as in Fig. \ref {fig:MS_NaO}, adopted from \citet{carretta09}. Dilution factors ranging from 1 to 100 are shown for each model (assorted symbols) and connected for each f value (coloured lines).}
    \label{fig:rotatingdilution}
\end{figure}
 \begin{figure}
    \centering
    \includegraphics[width = \columnwidth]{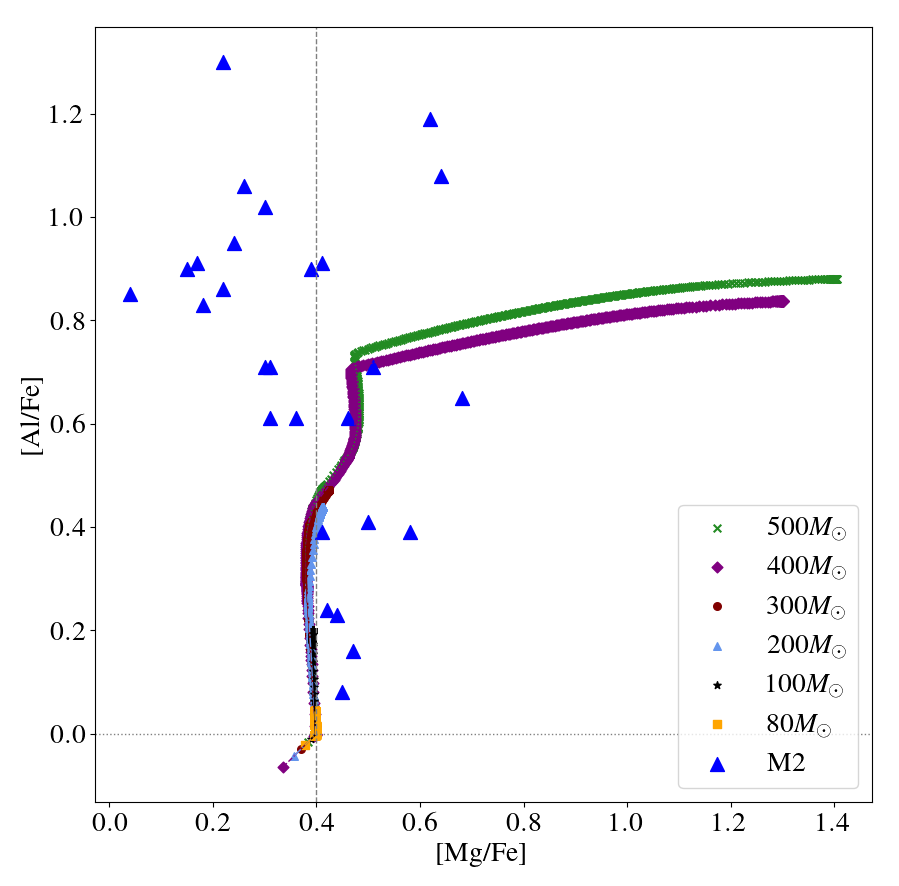}
    \caption{Cumulative ejected masses of [Mg/Fe] and [Al/Fe] for 80-500\Mdot\ models calculated at 1/30th \Zdot. Observations of stars in M2 are shown by blue triangles. The [Mg/Fe] ratio includes isotopes of $^{24,25,26}$Mg and $^{26}$Al, while the [Al/Fe] ratio is calculated for $^{27}$Al.}
    \label{fig:MgAl}
\end{figure}

Finally, we showcase the ejecta from our models alongside observed abundance patterns to directly compare the composition of P2 stars with the composition of the ejected mass from 80-500\Mdot\ stars with the same [Fe/H] metallicity. Figure \ref{fig:GC_EM_hist} presents the cumulative ejected masses integrated from the onset of core H-burning until core He-exhaustion, by adopting equation \ref{EMeq} and integrating over short timescales covered in 10 model steps throughout the entire evolution. After calculating the ejected mass of Na and O for each segment, the ejecta are normalised to the [Na/Fe] and [O/Fe] notation, by scaling the log (Na/Fe) fractions by number relative to solar values. Each integrated ejected mass is presented by a marker, which is different for each initial mass model, and then connected by a dashed line to show the evolution of these cumulative ejected masses over time as they become increasingly Na-rich and O-depleted. The same observations from NGC 5904 shown in Figs. \ref{fig:MS_NaO} and \ref{fig:postMS_NaO} are included, where the green ellipse encompasses P1 stars which have a surface composition representative of field stars, while those lying beyond the green ellipse are considered P2 stars. Figure \ref{fig:GC_EM_hist} confirms that all models begin their evolution by ejecting material with a composition (relative to solar) similar to P1 stars. Then, as the higher mass models begin to peel their outer layers to expose fresh nucleosynthesised material, the ejecta quickly becomes Na-rich and O-poor. This behaviour occurs for all masses in the range 80-500\Mdot\ but as previously discussed, happens later in the evolution for 80-100\Mdot\ models than for 200-500\Mdot\ models. Therefore, the longest evolutionary phase with the highest mass loss, which provides ejected masses with abundance patterns which are representative of the P2 stars, is the MS evolution of VMS with masses $\sim$ 200\Mdot. 

A key point in establishing the second population in GCs is that the ejected material should be slow-moving such that it remains in the potential well of the cluster to form a second population of stars. Supernovae ejecta and WR winds experience velocities of a few thousand km/s which are much higher than the predicted escape velocities of GCs \citep[e.g.][]{Gieles18}, while VMS winds are likely slower than those from WR stars \citep{vink18,Vink23}. In fact, since VMS can eject synthesised material from the onset of core H-burning for a few million years, and the second population of stars in GCs can be formed within a few million years of the first population \citep{Gratton12}, VMS could in fact be an important source for forming P2 stars in GCs. Moreover, \cite{Gratton12} discusses the observed spread in Fe abundances in Galactic GCs which resembles similar patterns observed in local dwarf galaxies and suggests that globular clusters may start their evolution as the cores of dwarf galaxies \citep{bekki03, bekki07, boker08, carretta10}. Interestingly, a large homogeneous sample of over 900 massive stars in the 30 Doradus region of the Large Magellanic Cloud was investigated by the VLT-FLAMES Tarantula Survey \citep[VFTS;][]{Evans11}, exploring the complex star-formation of the nebula, as well as detecting the most massive stars to date with M$_{*}$ $\sim$ 200\Mdot. Furthermore, \cite{Schneider18} found an excess of massive stars in the region and a top-heavy IMF with a power law index of 1.90. These recent studies collectively adhere to the theory that VMS may be the most parsimonious solution to self-enrichment in GCs \citep{vink18}.

\section{Discussion}
\subsection{Dilution}\label{discussion_dilute}
The extent of the Na-O anti-correlation observed across numerous clusters has provided important clues for constraining the multiple population phenomenon in GCs. While some ejecta from the first population may be encapsulated to form the second population, it is likely that a fraction of the mass accumulated will be from the original cluster gas. Therefore, we test the effect of diluting our wind ejecta with the initial composition to estimate the dilution factor required to reproduce the observed abundance ratios of Na and O. We calculate the dilution as
\begin{equation}
  X_{D} =   X_{i} \times \frac{f-1}{f} + \frac{M_{EM}}{\Delta M} \times \frac{1}{f}
\end{equation}
where $X_{D}$ is the abundance of the diluted ejecta, $X_{i}$ is the initial abundance of the original gas for a given isotope, $f$ is the dilution factor, $M_{EM}$ is the total wind ejecta of a given isotope and $\Delta$ $M$ represents the total mass lost. Figure \ref{fig:dilution} shows the total cumulative ejecta of [Na/Fe] and [O/Fe] for models with initial masses ranging from 80-300\Mdot\ since stars below 80\Mdot\ do not eject a meaningful amount of mass, and stars beyond 300\Mdot\ eject too much O-rich material to provide a total ejecta representative of the O-depleted population. We dilute the ejecta by factors ranging from 1 to 100, and show the resulting diluted gas abundance for each f value with a different coloured line. We find that we require a low dilution factor of f$=$ 2-4 in order to best represent the observed [Na/Fe] abundance pattern, since VMS with M$=$ 100-300\Mdot\ can reach the increased [Na/Fe] values of up to 0.8dex. The [O/Fe] ratio is more challenging to explain with diluted values. However, the undiluted VMS models can reach the negative [O/Fe] values thus possibly explaining the full range of observed abundances.

We include a subset of rotating models calculated at 1/30th \Zdot, with $\Omega_{\rm{ini}} / \Omega_{\rm{crit}}$ $=$ 0.4, and note that some models do not complete core He-burning since higher rotation rates lead to stars approaching criticality at such low Z. The impact of rotation on the observed [Na/Fe]-[O/Fe] correlation is shown in Fig. \ref{fig:rotatingdilution} for a range of dilution factors, comparable to Fig. \ref{fig:dilution}. We find that the 80\Mdot\ model does not lose significant mass on the main sequence, but loses significant amounts of $^{16}$O during core He-burning which results in O-production rather than depletion, regardless of dilution factor. The 80\Mdot\ rotating model reaches the cWR phase at effective temperatures ranging from log $T_{\rm{eff}}$ $=$ 5.0-5.2, leading to 0.1\Mdot\ of $^{16}$O ejecta (see Table \ref{Z0006Omega04_yields}). This is in contrast to the non-rotating models shown in Fig. \ref{fig:dilution} which displays a trend of increasing Na-enrichment and O-depletion. The 100-300\Mdot\ models are not affected in the same way since their convective cores are almost the entire stellar mass, leading to negligible effects of rotational mixing, and therefore minor changes in Na production and O destruction. Figure \ref{fig:rotatingdilution} illustrates that Na is overproduced in all rotating models, and by diluting the material, the [Na/Fe] shifts towards lower values until reaching the initial composition representative of field stars with f$=$ 100. On the other hand, the difference between O depletion by the 100-300\Mdot\ models and O production by the 80\Mdot\ model leads to 2 paths of diluting back to the initial composition, one which ranges from $\sim$ -1.0 up to 0.4 (similar to the behaviour seen in Fig. \ref{fig:dilution}), or from 1.5 down to the original gas value of 0.4, respectively.

\subsection{Mg-Al anti-correlation}
The Na-O anti-correlation observed in GCs has been the focus of many studies \citep[e.g.][]{carretta09,carretta10}, with many progenitors explored to reproduce the observed abundance patterns and ultimately explain the formation of multiple populations. Another indicator of the second population in GCs is the observed Mg-Al anti-correlation, which has not been observed in all clusters, and in many cases is not as extended or evident in less massive or metal-rich clusters. In some sense, it is clear that when considering the Na-O anti-correlation, we expect Na to enrich with nucleosynthesis and O to be depleted (at least in the early stages of nuclear burning where most of the stellar lifetime is spent). In comparison, the Mg-Al cycle replenishes $^{24}$Mg while also forming Al, and therefore, it is less evident how such Mg-depleted stars could have such Al-enrichment. In Fig. \ref{fig:MgAl} we present our non-rotating model grid calculated at 1/30th \Zdot\ alongside observations from NGC 7089 (M2) which has a similar [Fe/H] abundance (–1.47dex $\pm$ 0.03) and a broad range of Mg and Al. We find that our models (scaled for $\alpha$-enhancement) can represent the wide range of Al-enhancement observed, when considering the highest mass models. On the other hand, even when including all isotopes which contribute to observed Mg ($^{24, 25, 26}$Mg and $^{26}$Al which decays to $^{26}$Mg), the models do not become sufficiently Mg-depleted to enclose the observed data. \cite{Decressin07} suggested that the $^{24}$Mg(p, $\gamma$) reaction would need to be increased by a factor of 1000 to reproduce the observed Mg-depletion. Yet while many experimental rates suggest an upper/lower limit corresponding to $\sim$20\% uncertainty in this rate, it is unlikely that the reaction would be revised to such an extreme case. Moreover, \cite{choplin} suggests a similar increase by a factor 100 in the $^{23}$Na(p, $\gamma$)$^{24}$Mg reaction, yet a significant change to this reaction rate would also effect the Na-enrichment observed in GCs due to the depletion of $^{23}$Na at the expense of $^{24}$Mg.

\subsection{Wind yields at low Z}
The contribution of nuclear-processed elements from stellar winds at low Z is considered to be diminished due to the effect of the metallicity dependence on radiatively-driven winds. This may be the case for canonical O stars with initial masses ranging from 20-80\Mdot\ which drive optically thin winds with mass-loss rates of order 10$^{-6}$ to 10$^{-8}$. However, rapid rotation can induce higher mass-loss rates as stars get closer to their $\Omega\Gamma$-limit, leading to increased winds in lower Z environments. In this scenario, massive stars would need to rotate at rapid rotation rates ($\Omega_{\rm{ini}} / \Omega_{\rm{crit}}$ > 0.4) in order to increase their wind rate sufficiently to impact their overall wind yields. We note that while the effect of rotational mixing at low metallicity is an interesting topic of stellar evolution, we do not investigate the full effects of rotation in this work. Since VMS are almost fully convective, with cores which are $\geq$ 90\% of the total stellar mass, the impact of rotational mixing at these high masses are significantly reduced, even at the lowest Z. We do however, test the effect of rotationally-enhanced winds due to the $\Omega\Gamma$-limit as a function of chemical wind yields, but note that at Z $<$ 10\% \Zdot, models with $\Omega_{\rm{ini}} / \Omega_{\rm{crit}}$ > 0.4 reach criticality early in the evolution. Furthermore, what occurs in Nature to resolve stars at criticality is undetermined, and potentially points to unresolved physics which is beyond the scope of this work. 

On the other hand, above $\sim$ 80\Mdot\ stars approach the transition point where more massive stars launch optically thick winds which lead to enhanced mass-loss rates of up to 10$^{-3}$ even at low Z. This means that while the base mass-loss rates are reduced with Z, VMS above the transition point can still be impactful in their wind ejecta. We find that at 10\% \Zdot\ stars with initial masses ranging from 200-500\Mdot\ still lose 60-90\% of their total mass, and at 1/30th \Zdot\ the most massive stars continue to lose up to 60\% of their total mass, ejecting hundreds of solar masses of nucleosynthesised material into the ISM (see Table \ref{tab:appZ0006EM}). This suggests that VMS maintain a key role in the enrichment of the early Universe, particularly in their pollution of H-processed elements such as $^{14}$N and $^{23}$Na.

\section{Conclusions}\label{conclusions}
In this work, we explore the effect of optically thick enhanced winds on the nucleosynthesis and chemical yields of massive and very massive stars across low metallicities. We provide a range of initial masses for 4 metallicities scaling from \ZSMC\ down to 1\% \Zdot. We test the implementation of rotationally enhanced mass loss due to the $\Omega\Gamma$ limit for \ZSMC\ and 10\% \Zdot\ models, and focus on non-rotating models for 3\% and 1\% \Zdot\ models. We provide ejected masses and net wind yields for all grids of models in Appendix \ref{tables}.

We find that due to the increasingly hot and compact evolution of stars with lower Z, the central temperature changes can have an effect on the nucleosynthesis of light elements from Na-Mg-Al isotopes. However, while the relative changes in these reactions and subsequent surface abundances may change by a few 0.1 dex, the mass-loss rates can change by orders of magnitude across this Z range. Therefore, the wind rates dominate the yields of a given isotope rather than the effect of host Z environment on the nucleosynthesis.

Rotating (70\% critical) models can lose up to 100\Mdot\ more than their non-rotating counterparts during the MS. The effects of rotation are diminished at the highest mass range due to the loss of angular momentum early in the evolution. Counter-intuitively, the indirect effects of mass loss on rotation, while also considering the effects of rotationally-induced mass loss, can mean that intermediate rotation rates (40\% critical) can lose the most mass, spinning up at low Z towards critical, and result in higher ejecta masses than higher or lower rotation rates.

A Na-O anti-correlation is produced at the surface of VMS stars (M$>$100\Mdot) during the MS as well as at the surface of 80-100\Mdot\ stars during the post-MS phase. Furthermore, the substantial amounts of Na-rich and O-depleted material lost by 200-300\Mdot\ mass stars on their MS suggests that this phase would be most effective in enriching the ISM to form a second population of stars. We compare the normalised ejected masses of our 3\% \Zdot\ models with observations of red giant stars from the globular cluster NGC 5904. We find that our models naturally produce Na-enriched and O-depleted material, without tailoring model inputs to fit the observations, qualitatively reproducing the observed trend of [Na/Fe] and [O/Fe] with the surface abundances of the second population.

 \begin{acknowledgements}
The authors acknowledge MESA authors and developers for their continued revisions and public accessibility of the code. JSV, AML, and ERH are supported by STFC funding under grant number ST/V000233/1 in the context of the BRIDGCE UK Network. RH acknowledges support from STFC, the World Premier International Research Centre Initiative (WPI Initiative), MEXT, Japan and the IReNA AccelNet Network of Networks (National Science Foundation, Grant No. OISE-1927130). This article is based upon work from the ChETEC COST Action (CA16117) and the European Union’s Horizon 2020 research and innovation programme (ChETEC-INFRA, Grant No. 101008324).
 \end{acknowledgements}

\section*{Data Availability}
The data underlying this article will be shared on reasonable request to the corresponding author.
\typeout{}

\bibliographystyle{aa}
\bibliography{newdiff2.bib}

\begin{thebibliography}{58}
\expandafter\ifx\csname natexlab\endcsname\relax\def\natexlab#1{#1}\fi

\bibitem[{{Anders} \& {Grevesse}(1989)}]{AG89}
{Anders}, E. \& {Grevesse}, N. 1989, \gca, 53, 197

\bibitem[{{Arnett} {et~al.}(2019){Arnett}, {Meakin}, {Hirschi}, {Cristini}, {Georgy}, {Campbell}, {Scott}, {Kaiser}, {Viallet}, \& {Moc{\'a}k}}]{arnett19}
{Arnett}, W.~D., {Meakin}, C., {Hirschi}, R., {et~al.} 2019, \apj, 882, 18

\bibitem[{{Asplund} {et~al.}(2009){Asplund}, {Grevesse}, {Sauval}, \& {Scott}}]{asplund09}
{Asplund}, M., {Grevesse}, N., {Sauval}, A.~J., \& {Scott}, P. 2009, \araa, 47, 481

\bibitem[{{Baraffe} {et~al.}(2001){Baraffe}, {Heger}, \& {Woosley}}]{Baraffe2001}
{Baraffe}, I., {Heger}, A., \& {Woosley}, S.~E. 2001, \apj, 550, 890

\bibitem[{{Bastian} \& {Lardo}(2018)}]{bastianlardo18}
{Bastian}, N. \& {Lardo}, C. 2018, \araa, 56, 83

\bibitem[{{Bekki} {et~al.}(2007){Bekki}, {Campbell}, {Lattanzio}, \& {Norris}}]{bekki07}
{Bekki}, K., {Campbell}, S.~W., {Lattanzio}, J.~C., \& {Norris}, J.~E. 2007, \mnras, 377, 335

\bibitem[{{Bekki} \& {Freeman}(2003)}]{bekki03}
{Bekki}, K. \& {Freeman}, K.~C. 2003, \mnras, 346, L11

\bibitem[{{Boeltzig} {et~al.}(2019){Boeltzig}, {Best}, {Pantaleo}, {Imbriani}, {Junker}, {Aliotta}, {Balibrea-Correa}, {Bemmerer}, {Broggini}, {Bruno}, {Buompane}, {Caciolli}, {Cavanna}, {Chillery}, {Ciani}, {Corvisiero}, {Csedreki}, {Davinson}, {deBoer}, {Depalo}, {Di Leva}, {Elekes}, {Ferraro}, {Fiore}, {Formicola}, {F{\"u}l{\"o}p}, {Gervino}, {Guglielmetti}, {Gustavino}, {Gy{\"u}rky}, {Kochanek}, {Lugaro}, {Marigo}, {Menegazzo}, {Mossa}, {Munnik}, {Paticchio}, {Perrino}, {Piatti}, {Prati}, {Schiavulli}, {St{\"o}ckel}, {Straniero}, {Strieder}, {Sz{\"u}cs}, {Tak{\'a}cs}, {Trezzi}, {Wiescher}, \& {Zavatarelli}}]{Boeltzig}
{Boeltzig}, A., {Best}, A., {Pantaleo}, F.~R., {et~al.} 2019, Physics Letters B, 795, 122

\bibitem[{{B{\"o}ker}(2008)}]{boker08}
{B{\"o}ker}, T. 2008, \apjl, 672, L111

\bibitem[{{Bowman}(2020)}]{bowman20review}
{Bowman}, D.~M. 2020, Frontiers in Astronomy and Space Sciences, 7, 70

\bibitem[{{Bromm} \& {Larson}(2004)}]{Bromm2004}
{Bromm}, V. \& {Larson}, R.~B. 2004, \araa, 42, 79

\bibitem[{{Brott} {et~al.}(2011){Brott}, {de Mink}, {Cantiello}, {Langer}, {de Koter}, {Evans}, {Hunter}, {Trundle}, \& {Vink}}]{Brott2011}
{Brott}, I., {de Mink}, S.~E., {Cantiello}, M., {et~al.} 2011, \aap, 530, A115

\bibitem[{{Carretta} {et~al.}(2010){Carretta}, {Bragaglia}, {Gratton}, {Lucatello}, {Bellazzini}, {Catanzaro}, {Leone}, {Momany}, {Piotto}, \& {D'Orazi}}]{carretta10}
{Carretta}, E., {Bragaglia}, A., {Gratton}, R.~G., {et~al.} 2010, \apjl, 714, L7

\bibitem[{{Carretta} {et~al.}(2009){Carretta}, {Bragaglia}, {Gratton}, {Lucatello}, {Catanzaro}, {Leone}, {Bellazzini}, {Claudi}, {D'Orazi}, {Momany}, {Ortolani}, {Pancino}, {Piotto}, {Recio-Blanco}, \& {Sabbi}}]{carretta09}
{Carretta}, E., {Bragaglia}, A., {Gratton}, R.~G., {et~al.} 2009, \aap, 505, 117

\bibitem[{{Charbonnel}(2016)}]{charbonnel16}
{Charbonnel}, C. 2016, in EAS Publications Series, Vol. 80-81, EAS Publications Series, ed. E.~{Moraux}, Y.~{Lebreton}, \& C.~{Charbonnel}, 177--226

\bibitem[{{Choplin} {et~al.}(2018){Choplin}, {Meynet}, {Maeder}, {Hirschi}, \& {Chiappini}}]{choplin}
{Choplin}, A., {Meynet}, G., {Maeder}, A., {Hirschi}, R., \& {Chiappini}, C. 2018, in Journal of Physics Conference Series, Vol. 940, Journal of Physics Conference Series (IOP), 012021

\bibitem[{{Cristini} {et~al.}(2019){Cristini}, {Hirschi}, {Meakin}, {Arnett}, {Georgy}, \& {Walkington}}]{Cristini2019}
{Cristini}, A., {Hirschi}, R., {Meakin}, C., {et~al.} 2019, \mnras, 484, 4645

\bibitem[{{Cristini} {et~al.}(2017){Cristini}, {Meakin}, {Hirschi}, {Arnett}, {Georgy}, {Viallet}, \& {Walkington}}]{Cristini2017}
{Cristini}, A., {Meakin}, C., {Hirschi}, R., {et~al.} 2017, \mnras, 471, 279

\bibitem[{{Crowther} {et~al.}(2016){Crowther}, {Caballero-Nieves}, {Bostroem}, {Ma{\'\i}z Apell{\'a}niz}, {Schneider}, {Walborn}, {Angus}, {Brott}, {Bonanos}, {de Koter}, {de Mink}, {Evans}, {Gr{\"a}fener}, {Herrero}, {Howarth}, {Langer}, {Lennon}, {Puls}, {Sana}, \& {Vink}}]{crow16}
{Crowther}, P.~A., {Caballero-Nieves}, S.~M., {Bostroem}, K.~A., {et~al.} 2016, \mnras, 458, 624

\bibitem[{Cyburt {et~al.}(2010)Cyburt, Amthor, Ferguson, Meisel, Smith, Warren, Heger, Hoffman, Rauscher, Sakharuk, {et~al.}}]{Cyburt10}
Cyburt, R.~H., Amthor, A.~M., Ferguson, R., {et~al.} 2010, The Astrophysical Journal Supplement Series, 189, 240

\bibitem[{{de Jager} {et~al.}(1988){de Jager}, {Nieuwenhuijzen}, \& {van der Hucht}}]{deJager}
{de Jager}, C., {Nieuwenhuijzen}, H., \& {van der Hucht}, K.~A. 1988, \aaps, 72, 259

\bibitem[{{Decressin} {et~al.}(2007){Decressin}, {Meynet}, {Charbonnel}, {Prantzos}, \& {Ekstr{\"o}m}}]{Decressin07}
{Decressin}, T., {Meynet}, G., {Charbonnel}, C., {Prantzos}, N., \& {Ekstr{\"o}m}, S. 2007, \aap, 464, 1029

\bibitem[{{Denissenkov} \& {Hartwick}(2014)}]{Denissenkov}
{Denissenkov}, P.~A. \& {Hartwick}, F.~D.~A. 2014, \mnras, 437, L21

\bibitem[{{Ekstr{\"o}m} {et~al.}(2012){Ekstr{\"o}m}, {Georgy}, {Eggenberger}, {Meynet}, {Mowlavi}, {Wyttenbach}, {Granada}, {Decressin}, {Hirschi}, {Frischknecht}, {Charbonnel}, \& {Maeder}}]{Ekstroem2012}
{Ekstr{\"o}m}, S., {Georgy}, C., {Eggenberger}, P., {et~al.} 2012, \aap, 537, A146

\bibitem[{{Evans} {et~al.}(2011){Evans}, {Taylor}, {H{\'e}nault-Brunet}, {Sana}, {de Koter}, {Sim{\'o}n-D{\'{\i}}az}, {Carraro}, {Bagnoli}, {Bastian}, {Bestenlehner}, {Bonanos}, {Bressert}, {Brott}, {Campbell}, {Cantiello}, {Clark}, {Costa}, {Crowther}, {de Mink}, {Doran}, {Dufton}, {Dunstall}, {Friedrich}, {Garcia}, {Gieles}, {Gr{\"a}fener}, {Herrero}, {Howarth}, {Izzard}, {Langer}, {Lennon}, {Ma{\'{\i}}z Apell{\'a}niz}, {Markova}, {Najarro}, {Puls}, {Ramirez}, {Sab{\'{\i}}n-Sanjuli{\'a}n}, {Smartt}, {Stroud}, {van Loon}, {Vink}, \& {Walborn}}]{Evans11}
{Evans}, C.~J., {Taylor}, W.~D., {H{\'e}nault-Brunet}, V., {et~al.} 2011, \aap, 530, A108

\bibitem[{{Farmer} {et~al.}(2019){Farmer}, {Renzo}, {de Mink}, {Marchant}, \& {Justham}}]{Farmer+2019}
{Farmer}, R., {Renzo}, M., {de Mink}, S.~E., {Marchant}, P., \& {Justham}, S. 2019, \apj, 887, 53

\bibitem[{{Freytag} {et~al.}(1996){Freytag}, {Ludwig}, \& {Steffen}}]{Freytag1996}
{Freytag}, B., {Ludwig}, H.~G., \& {Steffen}, M. 1996, \aap, 313, 497

\bibitem[{{Gieles} {et~al.}(2018){Gieles}, {Charbonnel}, {Krause}, {H{\'e}nault-Brunet}, {Agertz}, {Lamers}, {Bastian}, {Gualandris}, {Zocchi}, \& {Petts}}]{Gieles18}
{Gieles}, M., {Charbonnel}, C., {Krause}, M. G.~H., {et~al.} 2018, \mnras, 478, 2461

\bibitem[{{Gratton} {et~al.}(2004){Gratton}, {Sneden}, \& {Carretta}}]{Gratt04}
{Gratton}, R., {Sneden}, C., \& {Carretta}, E. 2004, \araa, 42, 385

\bibitem[{{Gratton} {et~al.}(2012){Gratton}, {Carretta}, \& {Bragaglia}}]{Gratton12}
{Gratton}, R.~G., {Carretta}, E., \& {Bragaglia}, A. 2012, \aapr, 20, 50

\bibitem[{{Heger} {et~al.}(2000){Heger}, {Langer}, \& {Woosley}}]{Heger00}
{Heger}, A., {Langer}, N., \& {Woosley}, S.~E. 2000, \apj, 528, 368

\bibitem[{{Herwig}(2000)}]{Herwig2000}
{Herwig}, F. 2000, \aap, 360, 952

\bibitem[{{Higgins} \& {Vink}(2019)}]{Higgins}
{Higgins}, E.~R. \& {Vink}, J.~S. 2019, \aap, 622, A50

\bibitem[{{Higgins} {et~al.}(2023){Higgins}, {Vink}, {Hirschi}, {Laird}, \& {Sabhahit}}]{Higgins+23}
{Higgins}, E.~R., {Vink}, J.~S., {Hirschi}, R., {Laird}, A.~M., \& {Sabhahit}, G.~N. 2023, \mnras, 526, 534

\bibitem[{{Hirschi} {et~al.}(2005){Hirschi}, {Meynet}, \& {Maeder}}]{hirschi05}
{Hirschi}, R., {Meynet}, G., \& {Maeder}, A. 2005, \aap, 433, 1013

\bibitem[{{Krause} {et~al.}(2013){Krause}, {Fierlinger}, {Diehl}, {Burkert}, {Voss}, \& {Ziegler}}]{krause13}
{Krause}, M., {Fierlinger}, K., {Diehl}, R., {et~al.} 2013, \aap, 550, A49

\bibitem[{{Maeder} \& {Meynet}(2000)}]{MM00}
{Maeder}, A. \& {Meynet}, G. 2000, \aap, 361, 159

\bibitem[{{O'Connor} \& {Ott}(2011)}]{oconnor}
{O'Connor}, E. \& {Ott}, C.~D. 2011, \apj, 730, 70

\bibitem[{{Paxton} {et~al.}(2011){Paxton}, {Bildsten}, {Dotter}, {Herwig}, {Lesaffre}, \& {Timmes}}]{Pax11}
{Paxton}, B., {Bildsten}, L., {Dotter}, A., {et~al.} 2011, \apjs, 192, 3

\bibitem[{{Paxton} {et~al.}(2013){Paxton}, {Cantiello}, {Arras}, {Bildsten}, {Brown}, {Dotter}, {Mankovich}, {Montgomery}, {Stello}, {Timmes}, \& {Townsend}}]{Pax13}
{Paxton}, B., {Cantiello}, M., {Arras}, P., {et~al.} 2013, \apjs, 208, 4

\bibitem[{{Paxton} {et~al.}(2015){Paxton}, {Marchant}, {Schwab}, {Bauer}, {Bildsten}, {Cantiello}, {Dessart}, {Farmer}, {Hu}, {Langer}, {Townsend}, {Townsley}, \& {Timmes}}]{Pax15}
{Paxton}, B., {Marchant}, P., {Schwab}, J., {et~al.} 2015, \apjs, 220, 15

\bibitem[{{Paxton} {et~al.}(2018){Paxton}, {Schwab}, {Bauer}, {Bildsten}, {Blinnikov}, {Duffell}, {Farmer}, {Goldberg}, {Marchant}, {Sorokina}, {Thoul}, {Townsend}, \& {Timmes}}]{Pax18}
{Paxton}, B., {Schwab}, J., {Bauer}, E.~B., {et~al.} 2018, \apjs, 234, 34

\bibitem[{{Paxton} {et~al.}(2019){Paxton}, {Smolec}, {Schwab}, {Gautschy}, {Bildsten}, {Cantiello}, {Dotter}, {Farmer}, {Goldberg}, {Jermyn}, {Kanbur}, {Marchant}, {Thoul}, {Townsend}, {Wolf}, {Zhang}, \& {Timmes}}]{Pax19}
{Paxton}, B., {Smolec}, R., {Schwab}, J., {et~al.} 2019, \apjs, 243, 10

\bibitem[{{Pols} {et~al.}(1995){Pols}, {Tout}, {Eggleton}, \& {Han}}]{pols95}
{Pols}, O.~R., {Tout}, C.~A., {Eggleton}, P.~P., \& {Han}, Z. 1995, \mnras, 274, 964

\bibitem[{{Rizzuti} {et~al.}(2022){Rizzuti}, {Hirschi}, {Georgy}, {Arnett}, {Meakin}, \& {Murphy}}]{Rizzuti2022}
{Rizzuti}, F., {Hirschi}, R., {Georgy}, C., {et~al.} 2022, \mnras, 515, 4013

\bibitem[{Rogers \& Nayfonov(2002)}]{RogersNayfonov02}
Rogers, F. \& Nayfonov, A. 2002, The Astrophysical Journal, 576, 1064

\bibitem[{{Sabhahit} {et~al.}(2023){Sabhahit}, {Vink}, {Sander}, \& {Higgins}}]{sabh23}
{Sabhahit}, G.~N., {Vink}, J.~S., {Sander}, A. A.~C., \& {Higgins}, E.~R. 2023, \mnras, 524, 1529

\bibitem[{{Sander} \& {Vink}(2020)}]{SV2020}
{Sander}, A. A.~C. \& {Vink}, J.~S. 2020, \mnras, 499, 873

\bibitem[{{Schaerer} {et~al.}(2025){Schaerer}, {Guibert}, {Marques-Chaves}, \& {Martins}}]{Schaerer24}
{Schaerer}, D., {Guibert}, J., {Marques-Chaves}, R., \& {Martins}, F. 2025, \aap, 693, A271

\bibitem[{{Schneider} {et~al.}(2018){Schneider}, {Sana}, {Evans}, {Bestenlehner}, {Castro}, {Fossati}, {Gr{\"a}fener}, {Langer}, {Ram{\'\i}rez-Agudelo}, {Sab{\'\i}n-Sanjuli{\'a}n}, {Sim{\'o}n-D{\'\i}az}, {Tramper}, {Crowther}, {de Koter}, {de Mink}, {Dufton}, {Garcia}, {Gieles}, {H{\'e}nault-Brunet}, {Herrero}, {Izzard}, {Kalari}, {Lennon}, {Ma{\'\i}z Apell{\'a}niz}, {Markova}, {Najarro}, {Podsiadlowski}, {Puls}, {Taylor}, {van Loon}, {Vink}, \& {Norman}}]{Schneider18}
{Schneider}, F.~R.~N., {Sana}, H., {Evans}, C.~J., {et~al.} 2018, Science, 359, 69

\bibitem[{{Scott} {et~al.}(2021){Scott}, {Hirschi}, {Georgy}, {Arnett}, {Meakin}, {Kaiser}, {Ekstr{\"o}m}, \& {Yusof}}]{Scott21}
{Scott}, L.~J.~A., {Hirschi}, R., {Georgy}, C., {et~al.} 2021, \mnras, 503, 4208

\bibitem[{{Vink}(2018)}]{vink18}
{Vink}, J.~S. 2018, \aap, 615, A119

\bibitem[{{Vink}(2023)}]{Vink23}
{Vink}, J.~S. 2023, \aap, 679, L9

\bibitem[{{Vink} {et~al.}(2001){Vink}, {de Koter}, \& {Lamers}}]{Vink01}
{Vink}, J.~S., {de Koter}, A., \& {Lamers}, H.~J.~G.~L.~M. 2001, \aap, 369, 574

\bibitem[{{Vink} \& {Harries}(2017)}]{vink-harries17}
{Vink}, J.~S. \& {Harries}, T.~J. 2017, \aap, 603, A120

\bibitem[{{Vink} {et~al.}(2015){Vink}, {Heger}, {Krumholz}, {Puls}, {Banerjee}, {Castro}, {Chen}, {Chen{\`e}}, {Crowther}, {Daminelli}, {Gr{\"a}fener}, {Groh}, {Hamann}, {Heap}, {Herrero}, {Kaper}, {Najarro}, {Oskinova}, {Roman-Lopes}, {Rosen}, {Sander}, {Shirazi}, {Sugawara}, {Tramper}, {Vanbeveren}, {Voss}, {Wofford}, \& {Zhang}}]{vinkbook}
{Vink}, J.~S., {Heger}, A., {Krumholz}, M.~R., {et~al.} 2015, Highlights of Astronomy, 16, 51

\bibitem[{{Vink} {et~al.}(2011){Vink}, {Muijres}, {Anthonisse}, {de Koter}, {Gr{\"a}fener}, \& {Langer}}]{vink11}
{Vink}, J.~S., {Muijres}, L.~E., {Anthonisse}, B., {et~al.} 2011, \aap, 531, A132

\bibitem[{{Yusof} {et~al.}(2013){Yusof}, {Hirschi}, {Meynet}, {Crowther}, {Ekstr{\"o}m}, {Frischknecht}, {Georgy}, {Abu Kassim}, \& {Schnurr}}]{yusof13}
{Yusof}, N., {Hirschi}, R., {Meynet}, G., {et~al.} 2013, \mnras, 433, 1114

\end{thebibliography}
\begin{appendix}
\section{Additional figures}\label{figappendix}

\begin{figure}
    \centering
    \includegraphics[width = \columnwidth]{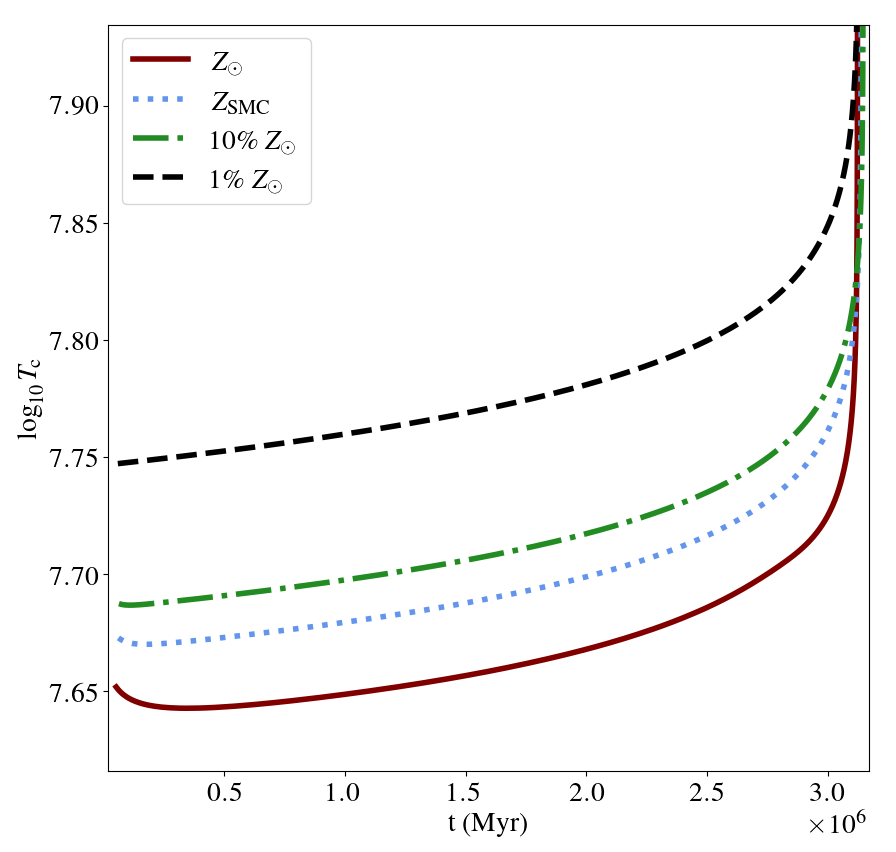}
    \caption{Central temperature evolution for 100\Mdot\ models during core H-burning calculated at various Z.}
    \label{fig:Tc_AllZ}
\end{figure}
\begin{figure}
    \centering
    \includegraphics[width = \columnwidth]{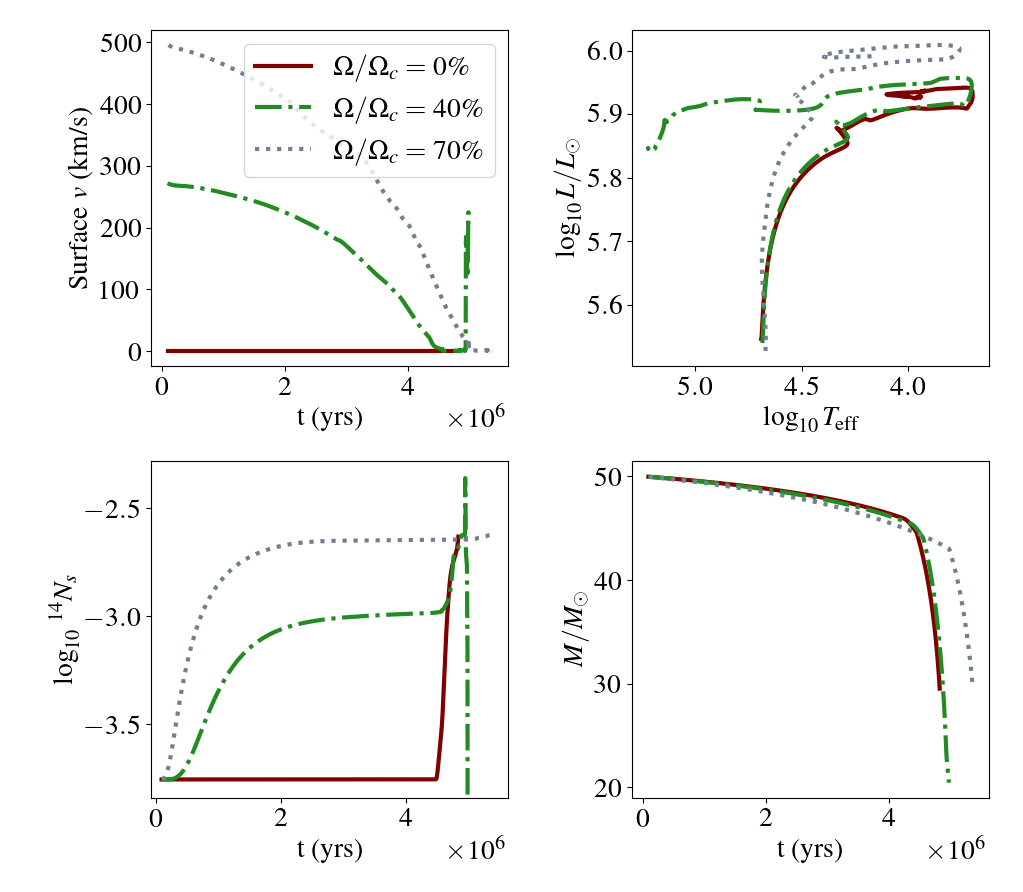}
    \caption{Surface rotation (top left), HRD (top right), surface $^{14}$N evolution (bottom left) and mass evolution (bottom right) of 50\Mdot\ models at \ZSMC\ for a range of critical rotation rates (0\% in solid red lines, 40\% in dash-dotted green lines and 70\% in dotted grey lines).}
    \label{fig:50M_SMC_panel}
\end{figure}
\begin{figure}
    \centering
    \includegraphics[width = \columnwidth]{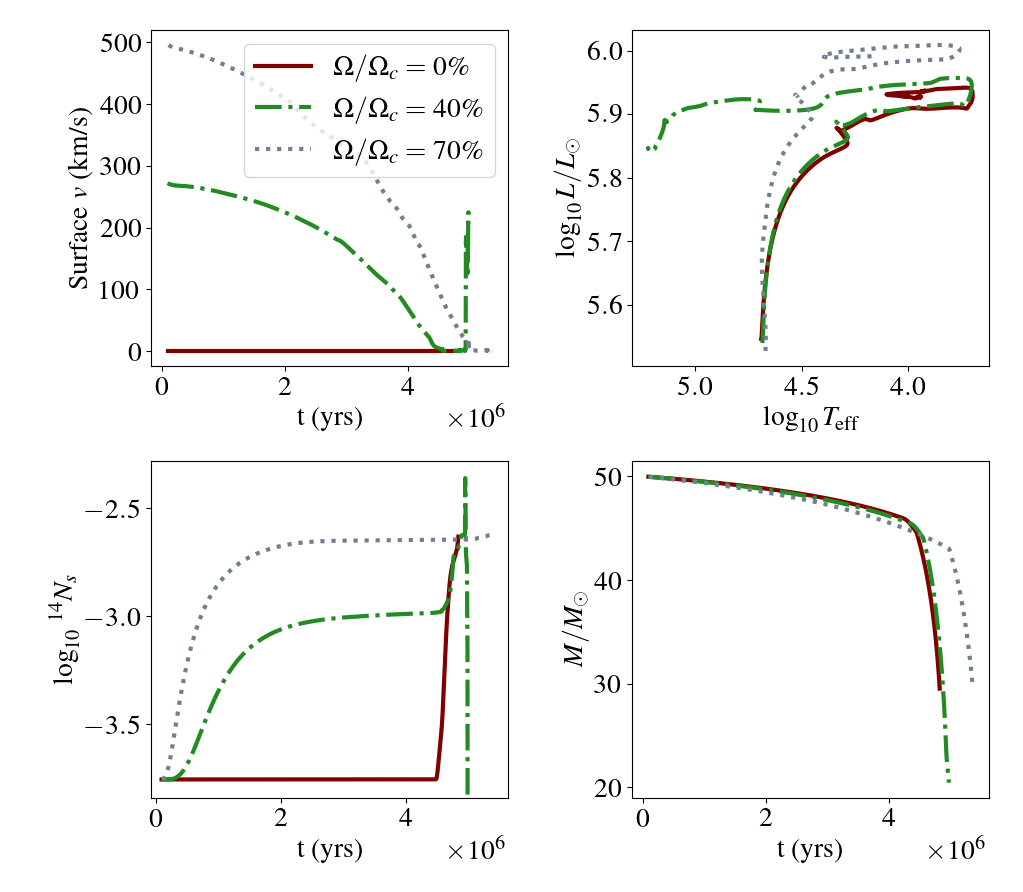}
    \caption{Surface rotation (top left), HRD (top right), surface $^{14}$N evolution (bottom left) and mass evolution (bottom right) of 50\Mdot\ models at \ZSMC\ for a range of critical rotation rates (0\% in solid red lines, 40\% in dash-dotted green lines and 70\% in dotted grey lines).}
    \label{fig:50M_SMC_panel}
\end{figure}
\begin{figure}
    \centering
    \includegraphics[width = \columnwidth]{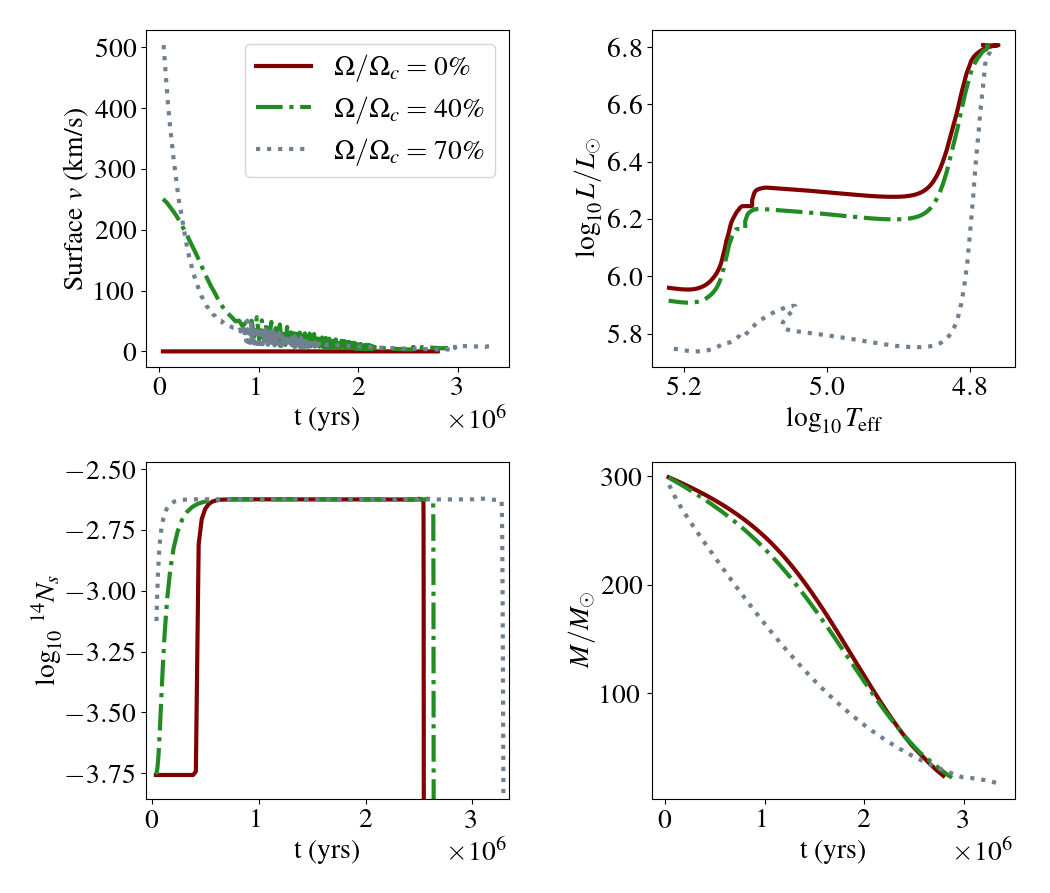}
    \caption{Surface rotation (top left), HRD (top right), surface $^{14}$N evolution (bottom left) and mass evolution (bottom right) of 300\Mdot\ models at \ZSMC\ for a range of critical rotation rates (0\% in solid red lines, 40\% in dash-dotted green lines and 70\% in dotted grey lines).}
    \label{fig:300M_SMC_panel}
\end{figure}

\section{Tables of ejected masses and net wind yields}\label{tables}
\onecolumn
\begin{landscape}
\begin{table*}

    \caption{Ejected masses (top) and net wind yields (bottom) in solar masses, for non-rotating (upper) 40\% critically rotating (middle), and 70\% critically rotating (bottom) stellar models at \ZSMC, calculated from the onset of core H-burning until core He-exhaustion.}
    \centering

\small
        \begin{tabular}{ccccccccccccccccc}

\hline
$M_{\rm{i}}/\rm{M}_{\odot}$ & $^{1}$H & $^{4}$He & $^{12}$C & $^{14}$N & $^{16}$O & $^{19}$F & $^{20}$Ne & $^{22}$Ne & $^{23}$Na & $^{24}$Mg & $^{25}$Mg & $^{26}$Mg & $^{26}$Al & $^{28}$Si \\
\hline \hline
50 & 1.27E+1 & 7.62 & 6.34E-3 & 2.08E-2 & 2.40E-2 & 1.74E-6 & 9.13E-3 & 4.95E-4 & 9.21E-4 & 2.86E-3 & 2.40E-4 & 4.10E-4 & 8.41E-5 & 3.40E-3 \\
80 & 2.13E+1 & 2.56E+1 & 5.59 & 6.90E-2 & 3.20 & 2.75E-6 & 2.76E-2 & 4.28E-2 & 4.86E-3 & 7.97E-3 & 2.86E-3 & 2.65E-3 & 3.40E-4 & 9.28E-3 \\
100 & 2.51E+1 & 3.41E+1 & 8.35 & 9.00E-2 & 4.80 & 3.64E-6 & 3.82E-2 & 6.65E-2 & 6.63E-3 & 1.04E-2 & 4.93E-3 & 4.61E-3 & 4.30E-4 & 1.20E-2 \\
200 & 5.18E+1 & 1.10E+2 & 8.10 & 3.21E-1 & 3.65 & 9.21E-6 & 7.43E-2 & 7.97E-2 & 1.99E-2 & 2.49E-2 & 3.51E-3 & 3.72E-3 & 1.75E-3 & 2.90E-2 \\
300 & 9.14E+1 & 1.71E+2 & 7.93 & 5.57E-1 & 3.55 & 9.38E-6 & 1.14E-1 & 7.84E-2 & 3.20E-2 & 3.90E-2 & 3.43E-3 & 4.10E-3 & 3.21E-3 & 4.57E-2 \\
400 & 1.31E+2 & 2.28E+2 & 8.83 & 7.83E-1 & 4.12 & 1.03E-5 & 1.55E-1 & 8.83E-2 & 4.33E-2 & 5.31E-2 & 4.22E-3 & 5.15E-3 & 4.64E-3 & 6.21E-2 \\
500 & 1.73E+2 & 2.84E+2 & 9.33 & 1.01 & 4.46 & 1.08E-5 & 1.96E-1 & 9.42E-2 & 5.44E-2 & 6.71E-2 & 4.74E-3 & 5.99E-3 & 6.12E-3 & 7.86E-2 \\
\hline
50 & -2.55 & 2.55 & -4.21E-3 & 1.72E-2 & -1.42E-2 & -1.13E-6 & -4.03E-4 & -2.76E-4 & 7.55E-4 & 3.27E-6 & -1.37E-4 & -2.17E-5 & 8.41E-5 & 9.79E-7 \\
80 & -2.05E+1 & 1.17E+1 & 5.56 & 5.92E-2 & 3.10 & -5.10E-6 & 1.54E-3 & 4.07E-2 & 4.41E-3 & 1.62E-4 & 1.83E-3 & 1.47E-3 & 3.40E-4 & -1.02E-9 \\
100 & -2.92E+1 & 1.61E+1 & 8.31 & 7.73E-2 & 4.66 & -6.56E-6 & 4.29E-3 & 6.38E-2 & 6.04E-3 & 2.46E-4 & 3.59E-3 & 3.07E-3 & 4.30E-4 & -8.14E-6 \\
200 & -7.88E+1 & 6.71E+1 & 8.01 & 2.91E-1 & 3.33 & -1.53E-5 & -7.03E-3 & 7.32E-2 & 1.85E-2 & 5.40E-4 & 2.87E-4 & 3.36E-5 & 1.75E-3 & 6.85E-5 \\
300 & -1.14E+2 & 1.03E+2 & 7.79 & 5.09E-1 & 3.04 & -2.91E-5 & -1.41E-2 & 6.81E-2 & 2.98E-2 & 7.96E-4 & -1.63E-3 & -1.70E-3 & 3.21E-3 & 1.11E-4 \\
400 & -1.48E+2 & 1.35E+2 & 8.64 & 7.18E-1 & 3.42 & -4.21E-5 & -1.92E-2 & 7.43E-2 & 4.03E-2 & 1.12E-3 & -2.66E-3 & -2.73E-3 & 4.64E-3 & 1.49E-4 \\
500 & -1.80E+2 & 1.67E+2 & 9.09 & 9.30E-1 & 3.58 & -5.55E-5 & -2.45E-2 & 7.64E-2 & 5.06E-2 & 1.45E-3 & -3.98E-3 & -3.99E-3 & 6.12E-3 & 1.86E-4 \\
\hline \hline    
50  & 1.43E+1 & 1.39E+1 & 2.93E-1 & 5.15E-2 & 9.32E-1 & 1.10E-6 & 1.99E-2 & 2.15E-3 & 2.31E-3 & 4.42E-3 & 1.85E-3 & 2.65E-3 & 2.11E-4 & 4.88E-3 \\
80  & 1.98E+1 & 2.66E+1 & 6.41 & 8.48E-2 & 3.56 & 1.67E-6 & 2.80E-2 & 4.89E-2 & 5.52E-3 & 8.07E-3 & 2.98E-3 & 2.90E-3 & 3.58E-4 & 9.39E-3 \\
100 & 2.35E+1 & 2.81E+1 & 5.39E-2 & 1.06E-1 & 2.14E-2 & 1.09E-6 & 2.17E-2 & 1.09E-3 & 4.83E-3 & 7.29E-3 & 1.80E-4 & 8.48E-4 & 4.56E-4 & 8.62E-3 \\
200 & 5.73E+1 & 1.12E+2 & 5.43 & 3.72E-1 & 2.11 & 5.72E-6 & 7.24E-2 & 5.20E-2 & 2.13E-2 & 2.52E-2 & 1.58E-3 & 2.33E-3 & 2.09E-3 & 2.95E-2 \\
300 & 9.72E+1 & 1.69E+2 & 6.49 & 5.92E-1 & 2.68 & 7.04E-6 & 1.13E-1 & 6.32E-2 & 3.26E-2 & 3.92E-2 & 2.27E-3 & 3.28E-3 & 3.48E-3 & 4.60E-2 \\
400 & 1.48E+2 & 2.20E+2 & 6.10 & 8.31E-1 & 2.47 & 6.89E-6 & 1.54E-1 & 5.98E-2 & 4.32E-2 & 5.32E-2 & 2.13E-3 & 3.77E-3 & 5.05E-3 & 6.28E-2 \\
500 & 1.91E+2 & 2.74E+2 & 6.93 & 1.06 & 2.94 & 7.84E-6 & 1.95E-1 & 6.88E-2 & 5.41E-2 & 6.72E-2 & 2.72E-3 & 4.62E-3 & 6.51E-3 & 7.93E-2 \\
\hline
50  & -7.78 & 6.56 & 2.78E-1 & 4.63E-2 & 8.76E-1 & -3.05E-6 & 6.15E-3 & 1.03E-3 & 2.07E-3 & 2.99E-4 & 1.31E-3 & 2.02E-3 & 2.11E-4 & -1.54E-5 \\
80  & -2.25E+1 & 1.26E+1 & 6.39 & 7.49E-2 & 3.46 & -6.26E-6 & 1.61E-3 & 4.68E-2 & 5.06E-3 & 1.77E-4 & 1.94E-3 & 1.71E-3 & 3.58E-4 & 1.37E-6 \\
100 & -1.53E+1 & 1.53E+1 & 2.72E-2 & 9.70E-2 & -7.54E-2 & -6.19E-6 & -2.46E-3 & -8.67E-4 & 4.41E-3 & 5.80E-5 & -7.77E-4 & -2.47E-4 & 4.56E-4 & 1.05E-5 \\
200 & -7.55E+1 & 6.80E+1 & 5.34 & 3.41E-1 & 1.78 & -1.92E-5 & -1.04E-2 & 4.53E-2 & 1.99E-2 & 4.31E-4 & -1.69E-3 & -1.42E-3 & 2.09E-3 & 7.13E-5 \\
300 & -1.10E+2 & 1.00E+2 & 6.34 & 5.43E-1 & 2.17 & -3.17E-5 & -1.58E-2 & 5.28E-2 & 3.03E-2 & 6.99E-4 & -2.83E-3 & -2.55E-3 & 3.48E-3 & 1.09E-4 \\
400 & -1.35E+2 & 1.26E+2 & 5.91 & 7.65E-1 & 1.77 & -4.61E-5 & -2.16E-2 & 4.56E-2 & 4.01E-2 & 8.79E-4 & -4.83E-3 & -4.20E-3 & 5.05E-3 & 1.37E-4 \\
500 & -1.66E+2 & 1.56E+2 & 6.68 & 9.72E-1 & 2.05 & -5.90E-5 & -2.68E-2 & 5.09E-2 & 5.02E-2 & 1.19E-3 & -6.08E-3 & -5.45E-3 & 6.51E-3 & 1.74E-4 \\
\hline \hline
50  & 1.02E+1 & 1.00E+1 & 9.93E-4 & 4.52E-2 & 2.89E-3 & 1.66E-7 & 8.63E-3 & 7.76E-5 & 1.82E-3 & 2.83E-3 & 3.42E-5 & 4.84E-4 & 1.26E-4 & 3.36E-3 \\
80  & 1.56E+1 & 2.72E+1 & 7.19 & 9.18E-2 & 4.62 & 7.19E-7 & 2.95E-2 & 5.39E-2 & 6.08E-3 & 7.89E-3 & 4.16E-3 & 3.93E-3 & 3.58E-4 & 9.10E-3 \\
100 & 1.83E+1 & 3.34E+1 & 8.10 & 1.10E-1 & 7.54 & 4.14E-7 & 4.77E-2 & 5.85E-2 & 7.50E-3 & 9.94E-3 & 9.27E-3 & 9.49E-3 & 4.77E-4 & 1.12E-2 \\
200 & 7.62E+1 & 1.03E+2 & 1.57 & 4.06E-1 & 4.58E-1 & 2.18E-6 & 7.43E-2 & 1.42E-2 & 2.02E-2 & 2.54E-2 & 3.95E-4 & 1.93E-3 & 2.38E-3 & 3.03E-2 \\
300 & 1.40E+2 & 1.42E+2 & 7.45E-1 & 6.39E-1 & 2.06E-1 & 2.06E-6 & 1.18E-1 & 7.48E-3 & 2.83E-2 & 3.88E-2 & 5.28E-4 & 3.09E-3 & 3.95E-3 & 4.71E-2 \\
400 & 9.80E+1 & 2.31E+2 & 1.69E+1 & 7.10E-1 & 1.06E+1 & 1.39E-5 & 1.73E-1 & 1.70E-1 & 4.15E-2 & 5.14E-2 & 1.55E-2 & 1.66E-2 & 4.10E-3 & 5.94E-2 \\
500 & 1.28E+2 & 2.91E+2 & 1.90E+1 & 9.13E-1 & 1.26E+1 & 1.50E-5 & 2.21E-1 & 1.90E-1 & 5.20E-2 & 6.52E-2 & 1.97E-2 & 2.19E-2 & 5.36E-3 & 7.51E-2 \\
\hline
50  & -5.00 & 5.00 & -9.46E-3 & 4.17E-2 & -3.49E-2 & -2.68E-6 & -8.11E-4 & -6.86E-4 & 1.65E-3 & 6.46E-6 & -3.40E-4 & 5.62E-5 & 1.26E-4 & 1.88E-6 \\
80  & -2.54E+1 & 1.36E+1 & 7.17 & 8.22E-2 & 4.52 & -6.98E-6 & 3.90E-3 & 5.18E-2 & 5.63E-3 & 2.30E-4 & 3.15E-3 & 2.77E-3 & 3.58E-4 & -5.00E-6 \\
100 & -3.23E+1 & 1.66E+1 & 8.06 & 9.79E-2 & 7.41 & -9.08E-6 & 1.62E-2 & 5.59E-2 & 6.95E-3 & 5.03E-4 & 8.02E-3 & 8.06E-3 & 4.77E-4 & -6.18E-5 \\
200 & -6.00E+1 & 5.80E+1 & 1.47 & 3.74E-1 & 1.18E-1 & -2.34E-5 & -1.06E-2 & 7.38E-3 & 1.87E-2 & 2.22E-4 & -2.96E-3 & -1.91E-3 & 2.38E-3 & 4.80E-5 \\
300 & -7.21E+1 & 7.13E+1 & 5.99E-1 & 5.89E-1 & -3.23E-1 & -3.77E-5 & -1.39E-2 & -3.21E-3 & 2.60E-2 & 2.47E-4 & -4.71E-3 & -2.90E-3 & 3.95E-3 & 5.60E-5 \\
400 & -1.68E+2 & 1.41E+2 & 1.67E+1 & 6.47E-1 & 9.93 & -3.62E-5 & 5.81E-3 & 1.56E-1 & 3.86E-2 & 1.68E-3 & 8.85E-3 & 9.02E-3 & 4.10E-3 & 7.63E-5 \\
500 & -2.08E+2 & 1.77E+2 & 1.88E+1 & 8.34E-1 & 1.17E+1 & -4.84E-5 & 9.94E-3 & 1.73E-1 & 4.83E-2 & 2.36E-3 & 1.14E-2 & 1.24E-2 & 5.36E-3 & 8.27E-5 \\
\hline

    \end{tabular}

    \label{tab:SMC_yields}

\end{table*}
\end{landscape}

\begin{landscape}
\begin{table}
    \caption{Ejected masses (top) and net wind yields (bottom) in solar masses, for non-rotating (upper) 40\% critically rotating (middle), and 70\% critically rotating (bottom) stellar models at 10\% \Zdot, calculated from the onset of core H-burning until core He-exhaustion.}
    \centering
\small
\begin{tabular}{ccccccccccccccc}
\hline
$M_{\rm{i}}/\rm{M}_{\odot}$ & $^{1}$H & $^{4}$He & $^{12}$C & $^{14}$N & $^{16}$O & $^{19}$F & $^{20}$Ne & $^{22}$Ne & $^{23}$Na & $^{24}$Mg & $^{25}$Mg & $^{26}$Mg & $^{26}$Al & $^{28}$Si \\
\hline \hline
30 & 1.64 & 5.31E-1 & 5.62E-4 & 1.90E-4 & 2.03E-3 & 1.53E-7 & 5.08E-4 & 4.11E-5 & 8.85E-6 & 1.52E-4 & 2.01E-5 & 2.30E-5 & 2.08E-23 & 1.81E-4 \\
50 & 4.92 & 1.61 & 1.66E-3 & 6.69E-4 & 6.04E-3 & 4.53E-7 & 1.52E-3 & 1.22E-4 & 3.11E-5 & 4.56E-4 & 5.97E-5 & 6.90E-5 & 4.06E-7 & 5.43E-4 \\
80 & 2.08E+1 & 2.21E+1 & 2.47 & 3.30E-2 & 2.46 & 1.08E-6 & 1.41E-2 & 9.31E-3 & 2.16E-3 & 3.46E-3 & 1.55E-3 & 1.33E-3 & 1.46E-4 & 3.99E-3 \\
100 & 2.45E+1 & 2.97E+1 & 5.76 & 4.30E-2 & 4.60 & 1.22E-6 & 2.14E-2 & 2.14E-2 & 2.93E-3 & 4.65E-3 & 3.00E-3 & 2.52E-3 & 1.85E-4 & 5.38E-3 \\
200 & 5.78E+1 & 1.03E+2 & 2.56E+1 & 1.40E-1 & 2.53E+1 & 2.02E-6 & 1.74E-1 & 9.58E-2 & 8.85E-3 & 1.87E-2 & 2.84E-2 & 3.20E-2 & 6.27E-4 & 1.76E-2 \\
300 & 7.37E+1 & 1.36E+2 & 3.19E+1 & 1.86E-1 & 3.52E+1 & 2.16E-6 & 3.00E-1 & 1.17E-1 & 1.14E-2 & 3.15E-2 & 4.29E-2 & 5.28E-2 & 8.69E-4 & 2.30E-2 \\
500 & 9.37E+1 & 2.44E+2 & 3.15E+1 & 3.20E-1 & 2.92E+1 & 7.81E-6 & 2.46E-1 & 1.49E-1 & 1.65E-2 & 3.43E-2 & 3.58E-2 & 4.04E-2 & 1.79E-3 & 3.33E-2 \\
\hline
30 & -1.26E-7 & 2.56E-9 & -1.11E-9 & 5.18E-11 & -3.53E-13 & 1.38E-15 & -8.58E-14 & -2.06E-14 & 1.28E-14 & -2.58E-14 & -3.40E-15 & -3.89E-15 & 2.08E-23 & -3.06E-14 \\
50 & -1.59E-2 & 1.59E-2 & -2.77E-5 & 9.65E-5 & -7.74E-5 & -5.91E-9 & -2.50E-6 & -1.54E-6 & 4.49E-6 & 2.17E-8 & -7.56E-7 & -1.30E-7 & 4.06E-7 & 7.18E-9 \\
80 & -1.54E+1 & 1.04E+1 & 2.46 & 2.88E-2 & 2.41 & -2.28E-6 & 2.94E-3 & 8.40E-3 & 1.96E-3 & 1.13E-4 & 1.11E-3 & 8.23E-4 & 1.46E-4 & 1.03E-5 \\
100 & -2.42E+1 & 1.39E+1 & 5.74 & 3.74E-2 & 4.54 & -3.31E-6 & 6.33E-3 & 2.02E-2 & 2.67E-3 & 1.32E-4 & 2.40E-3 & 1.83E-3 & 1.85E-4 & 1.76E-5 \\
200 & -1.02E+2 & 5.08E+1 & 2.55E+1 & 1.22E-1 & 2.51E+1 & -1.28E-5 & 1.25E-1 & 9.18E-2 & 7.99E-3 & 3.96E-3 & 2.64E-2 & 2.98E-2 & 6.27E-4 & -2.06E-6 \\
300 & -1.35E+2 & 6.79E+1 & 3.18E+1 & 1.62E-1 & 3.49E+1 & -1.73E-5 & 2.36E-1 & 1.11E-1 & 1.03E-2 & 1.22E-2 & 4.04E-2 & 4.99E-2 & 8.69E-4 & -1.43E-5 \\
500 & -2.07E+2 & 1.46E+2 & 3.14E+1 & 2.85E-1 & 2.88E+1 & -2.02E-5 & 1.53E-1 & 1.41E-1 & 1.48E-2 & 6.41E-3 & 3.21E-2 & 3.62E-2 & 1.79E-3 & 2.39E-4 \\
\hline\hline
30 & 1.85 & 6.35E-1 & 2.61E-4 & 1.23E-3 & 1.64E-3 & 1.20E-7 & 5.75E-4 & 3.23E-5 & 3.04E-5 & 1.74E-4 & 1.71E-5 & 3.12E-5 & 5.16E-7 & 2.07E-4 \\
50 & 5.54 & 2.10 & 9.57E-4 & 4.81E-3 & 3.72E-3 & 2.70E-7 & 1.74E-3 & 7.52E-5 & 1.54E-4 & 5.34E-4 & 3.87E-5 & 9.94E-5 & 5.80E-6 & 6.36E-4 \\
80 & 1.84E+1 & 2.29E+1 & 1.09 & 4.70E-2 & 2.67 & 4.09E-7 & 2.31E-2 & 3.67E-3 & 2.52E-3 & 3.65E-3 & 2.71E-3 & 3.00E-3 & 1.57E-4 & 3.75E-3 \\
100 & 2.06E+1 & 2.90E+1 & 4.77 & 5.17E-2 & 5.91 & 3.68E-7 & 2.84E-2 & 1.69E-2 & 3.35E-3 & 4.55E-3 & 4.54E-3 & 4.24E-3 & 1.90E-4 & 5.02E-3 \\
200 & 3.30E+1 & 5.87E+1 & 9.84 & 1.00E-1 & 1.88E+1 & 2.25E-7 & 1.48E-1 & 2.65E-2 & 6.47E-3 & 1.39E-2 & 2.08E-2 & 2.48E-2 & 4.33E-4 & 9.93E-3 \\
300 & 6.30E+1 & 1.67E+2 & 1.47E+1 & 2.45E-1 & 9.05 & 4.49E-6 & 6.93E-2 & 7.24E-2 & 1.46E-2 & 1.88E-2 & 7.69E-3 & 6.48E-3 & 1.20E-3 & 2.13E-2 \\
400 & 9.68E+1 & 2.29E+2 & 1.56E+1 & 3.55E-1 & 9.82 & 4.83E-6 & 9.11E-2 & 7.67E-2 & 2.03E-2 & 2.64E-2 & 8.66E-3 & 7.48E-3 & 1.80E-3 & 2.95E-2 \\
500 & 1.23E+2 & 2.88E+2 & 1.90E+1 & 4.47E-1 & 1.32E+1 & 5.26E-6 & 1.25E-1 & 9.31E-2 & 2.46E-2 & 3.36E-2 & 1.29E-2 & 1.19E-2 & 2.33E-3 & 3.72E-2 \\
\hline
30 & -2.74E-2 & 2.75E-2 & -3.81E-4 & 1.02E-3 & -6.87E-4 & -5.49E-8 & -4.76E-6 & -1.46E-5 & 2.03E-5 & 1.08E-8 & -5.92E-6 & 4.95E-6 & 5.16E-7 & 4.87E-9 \\
50 & -2.31E-1 & 2.31E-1 & -1.02E-3 & 4.14E-3 & -3.43E-3 & -2.67E-7 & -4.34E-5 & -6.92E-5 & 1.23E-4 & 1.92E-7 & -3.20E-5 & 1.85E-5 & 5.80E-6 & 7.41E-8 \\
80 & -1.57E+1 & 1.19E+1 & 1.08 & 4.31E-2 & 2.63 & -2.76E-6 & 1.26E-2 & 2.82E-3 & 2.33E-3 & 4.99E-4 & 2.29E-3 & 2.52E-3 & 1.57E-4 & -1.66E-6 \\
100 & -2.49E+1 & 1.43E+1 & 4.75 & 4.64E-2 & 5.85 & -3.87E-6 & 1.43E-2 & 1.58E-2 & 3.11E-3 & 3.25E-4 & 3.98E-3 & 3.61E-3 & 1.90E-4 & 2.83E-6 \\
200 & -5.80E+1 & 2.92E+1 & 9.81 & 8.95E-2 & 1.87E+1 & -8.24E-6 & 1.19E-1 & 2.43E-2 & 5.98E-3 & 5.51E-3 & 1.97E-2 & 2.35E-2 & 4.33E-4 & -7.71E-5 \\
300 & -1.29E+2 & 1.05E+2 & 1.47E+1 & 2.23E-1 & 8.81 & -1.34E-5 & 9.98E-3 & 6.76E-2 & 1.35E-2 & 1.04E-3 & 5.34E-3 & 3.80E-3 & 1.20E-3 & 1.80E-4 \\
400 & -1.69E+2 & 1.44E+2 & 1.55E+1 & 3.24E-1 & 9.49 & -1.99E-5 & 8.95E-3 & 7.01E-2 & 1.88E-2 & 1.81E-3 & 5.40E-3 & 3.76E-3 & 1.80E-3 & 2.45E-4 \\
500 & -2.12E+2 & 1.80E+2 & 1.89E+1 & 4.09E-1 & 1.28E+1 & -2.59E-5 & 2.11E-2 & 8.48E-2 & 2.28E-2 & 2.65E-3 & 8.79E-3 & 7.20E-3 & 2.33E-3 & 3.10E-4 \\
\hline\hline
30 & 2.73 & 1.55 & 1.22E-4 & 4.55E-3 & 5.42E-4 & 3.60E-8 & 9.22E-4 & 1.10E-5 & 1.77E-4 & 2.99E-4 & 5.74E-6 & 6.42E-5 & 4.05E-6 & 3.56E-4 \\
50 & 8.35 & 1.31E+1 & 3.54E-4 & 2.50E-2 & 6.41E-4 & 3.55E-8 & 4.07E-3 & 3.52E-5 & 1.54E-3 & 1.53E-3 & 5.82E-6 & 1.32E-4 & 7.02E-5 & 1.79E-3 \\
80 & 9.70 & 1.59E+1 & 5.16E-4 & 2.94E-2 & 1.02E-3 & 6.14E-8 & 4.77E-3 & 5.49E-5 & 1.91E-3 & 1.83E-3 & 9.59E-6 & 1.04E-4 & 1.05E-4 & 2.14E-3 \\
100 & 1.28E+1 & 2.37E+1 & 5.70E-3 & 5.59E-2 & 8.14E-2 & 7.73E-8 & 8.63E-3 & 1.05E-4 & 2.73E-3 & 2.86E-3 & 1.14E-4 & 2.50E-4 & 1.62E-4 & 3.06E-3 \\
200 & 6.02E+1 & 1.08E+2 & 2.88 & 1.92E-1 & 1.12 & 1.32E-6 & 3.32E-2 & 1.28E-2 & 1.20E-2 & 1.26E-2 & 4.95E-4 & 6.09E-4 & 9.46E-4 & 1.44E-2 \\
300 & 8.97E+1 & 1.67E+2 & 6.33 & 2.92E-1 & 2.90 & 2.90E-6 & 5.37E-2 & 3.06E-2 & 1.76E-2 & 1.99E-2 & 1.65E-3 & 1.42E-3 & 1.48E-3 & 2.23E-2 \\
\hline
30 & -5.02E-1 & 5.02E-1 & -9.83E-4 & 4.17E-3 & -3.46E-3 & -2.65E-7 & -7.65E-5 & -6.97E-5 & 1.60E-4 & 2.63E-7 & -3.38E-5 & 1.90E-5 & 4.05E-6 & 1.09E-7 \\
50 & -7.85 & 7.85 & -5.19E-3 & 2.31E-2 & -1.94E-2 & -1.47E-6 & -9.41E-4 & -3.70E-4 & 1.45E-3 & 3.02E-5 & -1.92E-4 & -9.49E-5 & 7.02E-5 & 5.62E-6 \\
80 & -9.62 & 9.62 & -6.10E-3 & 2.72E-2 & -2.29E-2 & -1.74E-6 & -1.20E-3 & -4.28E-4 & 1.81E-3 & 4.76E-5 & -2.27E-4 & -1.67E-4 & 1.05E-4 & 8.46E-6 \\
100 & -1.49E+1 & 1.48E+1 & -3.72E-3 & 5.26E-2 & 4.72E-2 & -2.48E-6 & 9.03E-5 & -5.84E-4 & 2.58E-3 & 2.97E-4 & -2.24E-4 & -1.37E-4 & 1.61E-4 & 1.43E-5 \\
200 & -6.99E+1 & 6.60E+1 & 2.84 & 1.77E-1 & 9.60E-1 & -1.08E-5 & -7.04E-3 & 9.57E-3 & 1.13E-2 & 6.31E-4 & -1.10E-3 & -1.21E-3 & 9.46E-4 & 8.79E-5 \\
300 & -1.11E+2 & 1.02E+2 & 6.27 & 2.50E-1 & 2.67 & -1.41E-5 & -8.51E-3 & 2.58E-2 & 1.62E-2 & 1.21E-3 & -7.42E-4 & -1.38E-3 & 1.41E-3 & 1.55E-4 \\
\hline
\end{tabular}

\end{table}
\end{landscape}

\begin{landscape}
\begin{table}
    \caption{Ejected masses (top) and net wind yields (bottom) in solar masses, for non-rotating stellar models at Z $=$ 0.00067, calculated from the onset of core H-burning until core H-exhaustion.}
    \centering
    \small
    \begin{tabular}{cccccccccccccccc}
    
    \hline
     $M_{\rm{i}}/\rm{M}_{\odot}$	&$^{1}$H&	$^{4}$He	&$^{12}$C	&$^{14}$N&	$^{16}$O&	$^{19}$F	&$^{20}$Ne&	$^{22}$Ne&	$^{23}$Na	&$^{24}$Mg&	$^{25}$Mg	&$^{26}$Mg	&$^{26}$Al&	$^{27}$Al&	$^{28}$Si\\
\hline \hline
30 & 3.84E-1 & 1.22E-1 & 4.38E-5 & 1.48E-5 & 1.58E-4 & 1.19E-8 & 3.95E-5 & 3.20E-6 & 6.89E-7 & 1.19E-5 & 1.57E-6 & 1.79E-6 & 0 & 9.29E-7 & 1.41E-5 \\
50 & 1.18 & 3.76E-1 & 1.35E-4 & 4.57E-5 & 4.88E-4 & 3.67E-8 & 1.22E-4 & 9.85E-6 & 2.12E-6 & 3.66E-5 & 4.82E-6 & 5.52E-6 & 1.71E-36 & 2.86E-6 &4.34E-5 \\
80 & 3.17 & 1.01 & 3.62E-4 & 1.23E-4 & 1.31E-3 & 9.83E-8 & 3.27E-4 & 2.64E-5 & 5.69E-6 & 9.80E-5 & 1.29E-5 & 1.48E-5 & 1.03E-24 & 7.67E-6 &1.16E-4 \\
100 & 5.29 & 1.69 & 6.04E-4 & 2.05E-4 & 2.19E-3 & 1.64E-7 & 5.45E-4 & 4.41E-5 & 9.51E-6 & 1.64E-4 & 2.16E-5 & 2.47E-5 & 1.51E-21 & 1.28E-5 &1.94E-4 \\
200 & 2.23E+1 & 8.25 & 1.56E-3 & 5.73E-3 & 5.51E-3 & 4.10E-7 & 2.24E-3 & 1.22E-4 & 2.83E-4 & 7.27E-4 & 5.52E-5 & 9.44E-5 & 1.07E-5 & 1.00E-4 &8.52E-4 \\
300 & 3.96E+1 & 1.94E+1 & 1.75E-3 & 1.68E-2 & 5.78E-3 & 4.25E-7 & 4.02E-3 & 1.59E-4 & 8.96E-4 & 1.48E-3 & 5.81E-5 & 1.34E-4 & 3.76E-5 & 2.72E-4 &1.65E-3 \\
400 & 5.35E+1 & 2.99E+1 & 1.91E-3 & 2.63E-2 & 6.00E-3 & 4.37E-7 & 5.56E-3 & 2.02E-4 & 1.34E-3 & 2.19E-3 & 6.09E-5 & 1.63E-4 & 6.47E-5 & 4.25E-4 &2.34E-3 \\
500 & 6.85E+1 & 4.30E+1 & 2.08E-3 & 3.72E-2 & 6.20E-3 & 4.48E-7 & 7.30E-3 & 2.57E-4 & 1.81E-3 & 3.06E-3 & 6.39E-5 & 1.93E-4 & 1.03E-4 & 6.08E-4 &3.13E-3 \\
\hline
30 & -6.26E-10 & 1.46E-15 & -1.57E-16 & 8.23E-18 & 2.03E-18 & -1.81E-23 & 4.94E-19 & -1.16E-20 & 7.18E-21 & -3.03E-20 & -4.15E-21 & -4.10E-21 & 0 & -1.79E-22 &  -2.03E-20 \\
50 & -7.73E-9 & 1.88E-12 & -3.69E-13 & 1.97E-14 & 1.31E-17 & 2.03E-19 & 8.25E-19 & -4.56E-20 & 2.53E-20 & -1.39E-19 & -1.13E-20 & -9.46E-21 & 1.71E-36 & -3.78E-21 &  -1.23E-19 \\
80 & -6.46E-8 & 2.16E-9 & -2.02E-10 & 9.13E-12 & -8.86E-14 & 2.39E-16 & -2.21E-14 & -2.47E-15 & 3.29E-16 & -6.64E-15 & -8.76E-16 & -1.00E-15 & 1.03E-24 & -5.20E-16 & -7.88E-15 \\
100 & -2.09E-7 & 1.43E-8 & -4.78E-9 & 2.17E-10 & -1.52E-13 & 6.38E-15 & -1.94E-14 & -8.78E-14 & 8.98E-14 & -5.84E-15 & -7.70E-16 & -8.81E-16 & 1.51E-21 & -4.57E-16 & -6.92E-15 \\
200 & -8.62E-1 & 8.63E-1 & -1.08E-3 & 4.83E-3 & -4.07E-3 & -3.10E-7 & -1.50E-4 & -7.16E-5 & 2.41E-4 & 9.38E-6 & -3.95E-5 & -1.39E-5 & 1.07E-5 & 4.43E-5 & 1.06E-6 \\
300 & -5.17 & 5.17 & -3.36E-3 & 1.50E-2 & -1.27E-2 & -9.63E-7 & -5.88E-4 & -2.14E-4 & 8.15E-4 & 9.89E-5 & -1.24E-4 & -7.53E-5 & 3.76E-5 & 1.63E-4 & 7.90E-6 \\
400 & -9.78 & 9.78 & -5.31E-3 & 2.38E-2 & -2.01E-2 & -1.53E-6 & -9.59E-4 & -3.26E-4 & 1.23E-3 & 2.35E-4 & -1.97E-4 & -1.32E-4 & 6.47E-5 & 2.71E-4 & 1.63E-5 \\
500 & -1.60E+1 & 1.61E+1 & -7.56E-3 & 3.39E-2 & -2.87E-2 & -2.17E-6 & -1.41E-3 & -4.47E-4 & 1.66E-3 & 4.46E-4 & -2.81E-4 & -2.02E-4 & 1.03E-4 & 4.04E-4 & 2.86E-5 \\
\hline
\label{tab:appZ0006EM}
\end{tabular}
\end{table}
\end{landscape}

\begin{landscape}
\begin{table}
    \caption{Ejected masses (top) and net wind yields (bottom) in solar masses, for non-rotating stellar models at Z $=$ 0.0002, calculated from the onset of core H-burning until core H-exhaustion.}
    \centering
    \small
    \begin{tabular}{cccccccccccccccc}
    
    \hline
     $M_{\rm{i}}/\rm{M}_{\odot}$	&$^{1}$H&	$^{4}$He	&$^{12}$C	&$^{14}$N&	$^{16}$O&	$^{19}$F	&$^{20}$Ne&	$^{22}$Ne&	$^{23}$Na	&$^{24}$Mg&	$^{25}$Mg	&$^{26}$Mg	&$^{26}$Al&	$^{27}$Al&	$^{28}$Si\\
\hline \hline
30 & 2.82E-1 & 8.93E-2 & 9.58E-6 & 3.25E-6 & 3.47E-5 & 2.61E-9 & 8.65E-6 & 7.00E-7 & 1.51E-7 & 2.60E-6 & 3.43E-7 & 3.92E-7 & 0 & 2.03E-7 &  3.08E-6 \\
50 & 1.99 & 7.04E-1 & 6.35E-5 & 4.30E-5 & 2.37E-4 & 1.78E-8 & 6.13E-5 & 4.83E-6 & 2.02E-6 & 1.92E-5 & 2.35E-6 & 2.73E-6 & 4.72E-8 & 2.06E-6 & 2.25E-5 \\
80 & 5.16 & 3.32 & 1.44E-4 & 3.80E-4 & 5.41E-4 & 4.04E-8 & 1.67E-4 & 1.24E-5 & 1.23E-5 & 5.95E-5 & 5.65E-6 & 6.88E-6 & 1.63E-6 & 1.74E-5 & 9.69E-5 \\
100 & 4.34 & 1.44 & 1.41E-4 & 8.44E-5 & 5.12E-4 & 3.84E-8 & 1.33E-4 & 1.04E-5 & 4.16E-6 & 4.15E-5 & 5.06E-6 & 5.86E-6 & 9.28E-8 & 3.61E-6 & 4.81E-5 \\
400 & 4.87E+1 & 6.82E+1 & 6.09 & 8.06 & 2.53E+1 & 6.03E-3 & 3.36E-1 & 1.19E+1 & 2.33E-1 & 1.94E-1 & 1.99 & 6.69 & 1.36E-4 & 1.12E-1 & 1.01E-1 \\
\hline
30 & -4.22E-10 & 4.35E-16 & -2.11E-17 & 1.15E-18 & -1.43E-19 & -6.55E-24 & -2.52E-20 & -1.14E-21 & -2.98E-22 & -7.75E-21 & -4.60E-22 & 6.81E-21 & 0 & 3.19E-21 & 4.70E-20 \\
50 & -5.71E-2 & 5.71E-2 & -5.94E-6 & 1.94E-5 & -1.46E-5 & -1.11E-9 & -1.33E-6 & -2.36E-7 & 9.31E-7 & 3.75E-7 & -1.36E-7 & -1.10E-7 & 4.72E-8 & 5.87E-7 & 2.19E-7 \\
80 & -1.28 & 1.28 & -7.46E-5 & 3.06E-4 & -2.51E-4 & -1.91E-8 & -3.10E-5 & -3.54E-6 & 8.90E-6 & 2.16E-7 & -2.17E-6 & -2.07E-6 & 1.63E-6 & 1.28E-5 & 2.66E-5 \\
100 & -5.32E-2 & 5.33E-2 & -8.17E-6 & 3.38E-5 & -2.81E-5 & -2.13E-9 & -2.07E-6 & -4.86E-7 & 1.81E-6 & 1.07E-6 & -2.77E-7 & -2.47E-7 & 9.28E-8 & 4.47E-7 &  9.87E-8 \\
400 & -8.80E+1 & 2.49E+1 & 6.09 & 8.06 & 2.52E+1 & 6.03E-3 & 3.31E-1 & 1.19E+1 & 2.33E-1 & 1.92E-1 & 1.99 & 6.69 & 1.36E-4 & 1.12E-1 & 9.97E-2 \\
\hline
\end{tabular}
\end{table}
\end{landscape}

\begin{landscape}
\begin{table}
    \caption{Ejected masses (top) and net wind yields (bottom) in solar masses, for 40 \% critically rotating stellar models at Z $=$ 0.00067, calculated from the onset of core H-burning until core H-exhaustion.}
    \centering
    \small
    \begin{tabular}{cccccccccccccccc}
    
    \hline
     $M_{\rm{i}}/\rm{M}_{\odot}$	&$^{1}$H&	$^{4}$He	&$^{12}$C	&$^{14}$N&	$^{16}$O&	$^{19}$F	&$^{20}$Ne&	$^{22}$Ne&	$^{23}$Na	&$^{24}$Mg&	$^{25}$Mg	&$^{26}$Mg	&$^{26}$Al&	$^{27}$Al&	$^{28}$Si\\
    \hline \hline
    30  & 1.02 & 3.67E-1 & 3.15E-5 & 2.96E-4 & 2.59E-4 & 1.94E-8 & 1.04E-4 & 5.18E-6 & 1.10E-5 & 3.26E-5 & 2.59E-6 & 5.59E-6 & 2.06E-7 & 3.45E-6 & 3.86E-5 \\
    50  & 4.11 & 3.00 & 1.65E-4 & 1.87E-3 & 8.97E-4 & 6.56E-8 & 4.96E-4 & 2.35E-5 & 8.67E-5 & 1.66E-4 & 1.65E-5 & 2.34E-5 & 7.06E-6 & 2.83E-5 & 1.99E-4 \\
    80  & 9.28 & 19.51 & 7.89E-2 & 1.13E-2 & 1.05E-1 & 4.31E-8 & 1.91E-3 & 1.56E-4 & 5.44E-4 & 6.15E-4 & 6.76E-5 & 5.67E-5 & 2.19E-4 & 2.88E-4 & 8.33E-4 \\
    100 & 8.93 & 19.25 & 2.36E-4 & 1.09E-2 & 6.79E-4 & 3.71E-8 & 1.66E-3 & 5.99E-5 & 5.35E-4 & 5.82E-4 & 4.56E-5 & 3.23E-5 & 2.31E-4 & 2.90E-4 & 8.11E-4 \\
    200 & 16.10 & 23.30 & 2.85E-4 & 1.52E-2 & 4.75E-4 & 3.02E-8 & 2.35E-3 & 5.88E-5 & 8.04E-4 & 1.22E-3 & 5.60E-6 & 2.37E-5 & 5.05E-5 & 2.55E-4 & 1.12E-3 \\
    500 & 67.88 & 107.72 & 4.61E-1 & 8.25E-1 & 8.11E-1 & 8.01E-8 & 1.16E-2 & 2.02E-3 & 2.78E-3 & 5.51E-3 & 4.36E-4 & 4.97E-4 & 6.81E-4 & 1.79E-3 & 5.22E-3 \\
\hline
    30  & -3.19E-2 & 3.20E-2 & -8.86E-5 & 2.55E-4 & -1.75E-4 & -1.33E-8 & -4.67E-6 & -3.59E-6 & 9.14E-6 & 3.65E-8 & -1.71E-6 & 6.72E-7 & 2.06E-7 & 9.03E-7 & 1.31E-8 \\
    50  & -1.28 & 1.28 & -4.49E-4 & 1.66E-3 & -1.33E-3 & -1.02E-7 & -5.90E-5 & -2.14E-5 & 7.70E-5 & -1.16E-7 & -5.45E-6 & -1.79E-6 & 7.06E-6 & 1.53E-5 & 1.65E-6 \\
    80  & -12.70 & 12.52 & 7.64E-2 & 1.04E-2 & 9.61E-2 & -6.38E-7 & -3.49E-4 & -2.68E-5 & 5.05E-4 & -6.41E-5 & -2.20E-5 & -4.59E-5 & 2.19E-4 & 2.35E-4 & 2.73E-5 \\
    100 & -12.45 & 12.45 & -2.20E-3 & 1.01E-2 & -8.14E-3 & -6.26E-7 & -5.42E-4 & -1.18E-4 & 4.96E-4 & -7.86E-5 & -4.15E-5 & -6.74E-5 & 2.31E-4 & 2.38E-4 & 2.68E-5 \\
    200 & -13.78 & 13.78 & -3.10E-3 & 1.39E-2 & -1.18E-2 & -8.91E-7 & -7.31E-4 & -1.89E-4 & 7.50E-4 & 2.99E-4 & -1.16E-4 & -1.16E-4 & 5.01E-5 & 1.83E-4 & 2.35E-5 \\
    500 & -66.94 & 64.78 & 4.46E-1 & 8.18E-1 & 7.57E-1 & -3.96E-6 & -2.24E-3 & 9.17E-4 & 2.51E-3 & 1.35E-3 & -1.04E-4 & -1.28E-4 & 6.71E-4 & 1.46E-3 & 2.72E-4 \\

\hline
\label{Z0006Omega04_yields}
\end{tabular}
\end{table}
\end{landscape}

\begin{table}
\centering
\small
\begin{tabular}{ccccccc}
\hline
$M_{\rm{i}}$	 & $M_{\rm{TAMS}}$	 & $M_{\rm{He core}}$	 & $M_{\rm{f}}$	 & $M_{\rm{CO core}}$	 & t$_{\rm{MS}}$ & t$_{\rm{He}}$ \\
& & & & & & \\
\hline \hline
\multicolumn{7}{c}{ $\Omega_{\rm{ini}} / \Omega_{\rm{crit}} = $ 0 } \\
\hline
30  & 28.66 & 11.17 & 26.84 & 9.71  & 6.50 & 0.54 \\
50  & 44.69 & 21.82 & 29.51 & 19.66 & 4.45 & 0.38 \\
80  & 63.05 & 38.63 & 24.03 & 20.85 & 3.46 & 0.33 \\
100 & 73.24 & 50.02 & 27.25 & 23.93 & 3.14 & 0.32 \\
200 & 54.24 & 51.77 & 24.53 & 21.32 & 2.61 & 0.32 \\
300 & 53.43 & 51.10 & 24.31 & 21.10 & 2.47 & 0.32 \\
400 & 57.68 & 55.62 & 25.51 & 22.24 & 2.37 & 0.32 \\
500 & 60.13 & 58.29 & 26.22 & 22.93 & 2.30 & 0.32 \\
\hline
\multicolumn{7}{c}{ $\Omega_{\rm{ini}} / \Omega_{\rm{crit}} = $ 0.4 } \\
\hline
30  & 28.48 & 11.33 & 26.48 & 9.93  & 6.52 & 0.55 \\
50  & 44.13 & 22.30 & 20.43 & 17.28 & 4.47 & 0.42 \\
80  & 62.91 & 39.57 & 23.37 & 20.16 & 3.48 & 0.33 \\
100 & 73.92 & 50.94 & 48.11 & 43.16 & 3.15 & 0.30 \\
300 & 46.88 & 44.35 & 22.51 & 19.39 & 2.51 & 0.33 \\
400 & 45.18 & 42.73 & 22.07 & 18.97 & 2.43 & 0.34 \\
500 & 48.82 & 46.59 & 23.03 & 19.89 & 2.34 & 0.33 \\
\hline
\multicolumn{7}{c}{ $\Omega_{\rm{ini}} / \Omega_{\rm{crit}} = $ 0.7 } \\
\hline
30  & 27.60 & 12.60 & 24.57 & 11.49 & 7.39 & 0.53 \\
50  & 43.05 & 23.77 & 29.70 & 22.13 & 4.87 & 0.39 \\
80  & 62.24 & 42.07 & 25.00 & 21.70 & 3.72 & 0.33 \\
100 & 75.06 & 53.82 & 32.24 & 28.70 & 3.35 & 0.31 \\
300 & 23.73 & 20.55 & 17.19 & 14.33 & 2.92 & 0.39 \\
400 & 102.51 & 101.25 & 40.47 & 36.52 & 2.24 & 0.29 \\
500 & 115.90 & 115.47 & 45.58 & 41.34 & 2.15 & 0.29 \\
\hline
\end{tabular}
\caption{Properties of Z$=$0.004 models for different initial velocities (v$=$0, v$=$0.4, v$=$0.7). Burning phases are presented in Myrs (t$_{\rm{MS}}$, t$_{\rm{He}}$), and all masses are shown in \Mdot. The final mass at core H-exhaustion ($M_{\rm{TAMS}}$) and core He-exhaustion ($M_{\rm{f}}$) are presented. Helium core masses are calculated at the end of core H-burning, and the CO core masses are provided at the end of core He-burning.}
\end{table}
\begin{table}
\centering
\begin{tabular}{ccccccc}
\hline
$M_{\rm{i}}$	 & $M_{\rm{TAMS}}$	 & $M_{\rm{He core}}$	 & $M_{\rm{f}}$	 & $M_{\rm{CO core}}$	 & t$_{\rm{MS}}$ & t$_{\rm{He}}$ \\
& & & & & & \\
\hline \hline
\multicolumn{7}{c}{ $\Omega_{\rm{ini}} / \Omega_{\rm{crit}} = $ 0 } \\
\hline
30  & 29.10 & 11.29 & 27.82 & 9.77  & 6.38 & 0.53 \\
50  & 46.91 & 22.17 & 43.44 & 19.40 & 4.37 & 0.38 \\
80  & 69.45 & 39.35 & 31.97 & 28.26 & 3.39 & 0.32 \\
100 & 82.30 & 51.02 & 35.28 & 31.42 & 3.08 & 0.31 \\
300 & 205.80 & 170.90 & 87.99 & 0.00  & 2.23 & 0.27 \\
400 & 269.90 & 231.50 & 122.50 & 0.00  & 2.11 & 0.26 \\
500 & 226.40 & 226.40 & 99.26  & 0.00  & 2.05 & 0.27 \\
\hline
\multicolumn{7}{c}{ $\Omega_{\rm{ini}} / \Omega_{\rm{crit}} = $ 0.4 } \\
\hline
30  & 28.94 & 11.44 & 27.51 & 10.06 & 6.59 & 0.55 \\
50  & 46.42 & 22.82 & 42.32 & 20.17 & 4.51 & 0.39 \\
80  & 69.75 & 40.67 & 34.75 & 30.82 & 3.51 & 0.34 \\
100 & 84.03 & 53.54 & 39.46 & 35.38 & 3.21 & 0.31 \\
200 & 154.40 & 116.20 & 79.13 & 0.00  & 2.55 & 0.27 \\
300 & 97.35 & 96.07 & 43.58 & 39.33 & 2.37 & 0.29 \\
400 & 102.30 & 101.80 & 45.36 & 41.01 & 2.26 & 0.29 \\
500 & 123.70 & 123.70 & 53.36 & 48.59 & 2.16 & 0.28 \\
\hline
\multicolumn{7}{c}{ $\Omega_{\rm{ini}} / \Omega_{\rm{crit}} = $ 0.7 } \\
\hline
30  & 28.05 & 13.22 & 25.71 & 12.42 & 7.74 & 0.51 \\
50  & 44.02 & 25.53 & 28.48 & 24.72 & 5.24 & 0.39 \\
80  & 65.81 & 46.65 & 54.30 & 45.92 & 3.95 & 0.32 \\
100 & 80.23 & 59.89 & 63.12 & 0.00  & 3.51 & 0.32 \\
200 & 39.41 & 36.68 & 26.51 & 23.05 & 2.86 & 0.33 \\
\hline
\end{tabular}
\caption{Properties of Z$=$0.002 models for different initial velocities (v$=$0, v$=$0.4, v$=$0.7). Burning phases are presented in Myrs (t$_{\rm{MS}}$, t$_{\rm{He}}$), and all masses are shown in \Mdot. The final mass at core H-exhaustion ($M_{\rm{TAMS}}$) and core He-exhaustion ($M_{\rm{f}}$) are presented. Helium core masses are calculated at the end of core H-burning, and the CO core masses are provided at the end of core He-burning.}
\end{table}
\begin{table}
\centering
\begin{tabular}{ccccccc}
\hline
$M_{\rm{i}}$	 & $M_{\rm{TAMS}}$	 & $M_{\rm{He core}}$	 & $M_{\rm{f}}$	 & $M_{\rm{CO core}}$	 & t$_{\rm{MS}}$ & t$_{\rm{He}}$ \\
\hline \hline
\multicolumn{7}{c}{Z$=$0.00067 } \\
\hline
30  & 29.49 & 11.37 & 28.75 & 9.77  & 6.60 & 0.73 \\
50  & 48.42 & 22.28 & 46.57 & 18.45 & 4.37 & 0.37 \\
80  & 75.79 & 39.85 & 71.57 & 30.72 & 3.39 & 0.31 \\
400 & 316.30 & 236.60 & 182.70 & 0.00  & 2.10 & 0.25 \\
500 & 388.20 & 298.50 & 222.00 & 0.00  & 2.02 & 0.25 \\
\hline
\multicolumn{7}{c}{Z$=$0.0002 } \\
\hline
30  & 29.78 & 11.40 & 29.63 & 9.72  & 6.37 & 0.52 \\
50  & 48.89 & 26.72 & 47.30 & 24.71 & 5.03 & 1.04 \\
80  & 75.84 & 51.29 & 71.50 & 0.00  & 3.95 & 0.86 \\
100 & 97.20 & 51.94 & 94.18 & 38.02 & 3.07 & 0.29 \\
\hline
\end{tabular}
\caption{Properties of non-rotating models calculated at Z$=$0.00067 and Z$=$0.0002. Burning phases are presented in Myrs (t$_{\rm{MS}}$, t$_{\rm{He}}$), and all masses are shown in \Mdot. The final mass at core H-exhaustion ($M_{\rm{TAMS}}$) and core He-exhaustion ($M_{\rm{f}}$) are presented. Helium core masses are calculated at the end of core H-burning, and the CO core masses are provided at the end of core He-burning.}
\end{table}

\begin{table}
\centering
\begin{tabular}{ccccccc}
\hline
$M_{\rm{i}}$	 & $M_{\rm{TAMS}}$	 & $M_{\rm{He core}}$	 & $M_{\rm{f}}$	 & $M_{\rm{CO core}}$	 & t$_{\rm{MS}}$ & t$_{\rm{He}}$ \\
   \hline \hline
    30  & 29.424 & 11.376 & 28.604 & 9.797 & 6.62E+06 & 5.44E+05 \\
    50  & 46.882 & 27.808 & 42.869 & 26.101 & 5.25E+06 & 1.08E+06 \\
    80  & 72.766 & 50.590 & 50.931 & 45.653 & 3.94E+06 & 5.58E+05 \\
    100 & 89.104 & 68.277 & 71.716 & 66.108 & 3.57E+06 & 5.41E+05 \\
    200 & 171.70 & 134.00 & 160.25 & 131.05 & 2.91E+06 & 2.56E+05\\
    500 & 406.656 & 309.748 & 321.144 & 290.32 & 2.06E+06 & 2.38E+05 \\
    \hline
    \end{tabular}
\caption{Properties of 40\% critically rotating models calculated at Z$=$0.00067. Burning phases are presented in Myrs (t$_{\rm{MS}}$, t$_{\rm{He}}$), and all masses are shown in \Mdot. The final mass at core H-exhaustion ($M_{\rm{TAMS}}$) and core He-exhaustion ($M_{\rm{f}}$) are presented. Helium core masses are calculated at the end of core H-burning, and the CO core masses are provided at the end of core He-burning.}
\end{table}

\end{appendix}
\end{document}